\documentclass{aa}
\usepackage{graphicx}
\graphicspath{{./figures/}}
\usepackage[varg]{txfonts}
\bibpunct{(}{)}{;}{a}{}{,} % to follow the A&A style
\usepackage{hyperref}
\usepackage{siunitx}

\newcommand{\diff}{\mathrm{d}}

\begin{document}

\title{Constructing stable 3D hydrodynamical models of giant stars}

\author{Sebastian~T.~Ohlmann\inst{1,2}\fnmsep\thanks{\email{sebastian.ohlmann@h-its.org}} \and
       Friedrich~K.~R\"{o}pke\inst{1,3} \and
       R\"{u}diger~Pakmor\inst{1} \and 
       Volker~Springel\inst{1,4}
       }

\titlerunning{Stable 3D hydro models of giant stars}
\authorrunning{S.~T.~Ohlmann et al.}

\institute{%
    Heidelberger Institut f\"{u}r Theoretische Studien,
    Schloss-Wolfsbrunnenweg 35, 69118 Heidelberg, Germany
  \and
    Institut f\"{u}r Theoretische Physik und Astrophysik,
    Universit\"{a}t W\"{u}rzburg, Emil-Fischer-Str. 31, 
    97074 W\"{u}rzburg, Germany
  \and
    Zentrum f\"ur Astronomie der Universit\"at Heidelberg,
    Institut f\"ur theoretische Astrophysik, Philosophenweg 12,
    69120 Heidelberg, Germany
  \and
    Zentrum f\"ur Astronomie der Universit\"at Heidelberg,
    Astronomisches Recheninstitut, M\"{o}nchhofstr. 12-14, 69120
    Heidelberg, Germany
}

\date{Received ; accepted }

\abstract
  {% Context
    Hydrodynamical simulations of stellar interactions require stable models of
    stars as initial conditions. Such initial models, however, are difficult to
    construct for giant stars because of the wide range in spatial
    scales of the hydrostatic equilibrium and in dynamical timescales between
    the core and the envelope of the giant.
    They are needed for, e.g., modeling the common envelope phase where
    a giant envelope encompasses both the giant core and a companion star.
  %} {% Aims
    Here, we present a new method of approximating and reconstructing giant
    profiles from a stellar evolution code to produce stable models for
    multi-dimensional hydrodynamical simulations.
  %} {% Methods
    We determine typical stellar stratification profiles with the
    one-dimensional stellar evolution code \textsc{mesa}\@. After an appropriate mapping,
    hydrodynamical simulations are conducted using the moving-mesh code
    \textsc{arepo}.
  %} {% Results
    The giant profiles are approximated by replacing the core of the giant with
    a point mass and by constructing a suitable continuation of the profile to
    the center. Different reconstruction methods are tested that can
    specifically control the convective behaviour of the model. After mapping to
    a grid, a relaxation procedure that includes damping of spurious velocities
    yields stable models in three-dimensional hydrodynamical simulations. 
    Initially convectively stable configurations lead to stable hydrodynamical
    models while for stratifications that are convectively unstable in the
    stellar evolution code, simulations recover the convective behaviour of the
    initial model and show large convective plumes with Mach numbers up to 0.8.
    Examples are shown for a $2M_\odot$ red giant and a $0.67M_\odot$
    asymptotic giant branch star.
  %} {% Conclusions 
    A detailed analysis shows that the improved method reliably provides stable
    models of giant envelopes that can be used as initial conditions for
    subsequent hydrodynamical simulations of stellar interactions involving giant
    stars.
  } 


\maketitle

\section{Introduction}
\label{sec:introduction}

Dynamical interactions of stars can be modelled with hydrodynamical simulations.
Most systems examined so far consist of rather compact stars, such as collisions
of main sequence (MS) stars as a model for blue stragglers
\citep[e.g.,][]{benz1987a,lombardi1996a,sills2001a}
or mergers of compact objects, i.e., white dwarfs (WDs), neutron stars (NSs), and
black holes \citep[see the review by][]{rosswog2015a}. These compact object
mergers may lead to a wide variety of phenomena, such as Type Ia supernovae from
merging WDs \citep[e.g.,][]{pakmor2013a}, or gravitational wave emission from
neutron star mergers \citep[e.g.,][]{bauswein2016a}.
An important interaction of more extended giant stars is the common envelope
(CE) phase. During this phase, the envelope of the primary giant star is shared
by the core of the giant and the companion star. Gravitational binding energy is
released and drives the ejection of the envelope. The result of this phase is
either a close binary or a merger of the stellar cores \citep[for a recent
review, see][]{ivanova2013a}.
The CE phase is needed to explain the evolution of stellar systems towards, e.g.,
close WD and MS binaries \citep{schreiber2003a,zorotovic2010a}, double WDs
\citep{han1995a,nelemans2001a}, and Type Ia supernovae
\citep{iben1984a,ruiter2009a,toonen2012a}.

To conduct hydrodynamical simulations of these stellar interactions, stable
models of stars are needed as initial conditions. 
In 1D stellar evolution approaches, the assumption of hydrostatic
equilibrium is usually part of the equations that are solved. For a 3D
hydrodynamics code, however, hydrostatic equilibrium is a special solution of
the more general equations that is not explicitely enforced; it can be easily
disturbed by mismatches between the discretizations of pressure and gravity.
For more compact stars, such as MS stars, WDs, or NSs, the representation of
their global hydrostatic structure causes
no particular problems because they span a small range of
temporal and spatial scales only. The difficulty in modeling WD or NS
interactions lies in the additional physics that is needed, as, e.g., a
degenerate equation of state, nuclear burning, or general relativity.
It is, however, harder to obtain stable models of giant stars because 
their hydrostatic equilibrium
encompass a wide range of scales in time and space and because their envelope is
usually only loosely bound. Thus, giant structures need to be approximated for
use in hydrodynamical simulations.

In recent years, numerical representations of giant structures have mainly been
used for modeling the dynamical spiral-in of a companion into a giant envelope
during a CE phase.  Hydrodynamical simulations use either a discretization on
computational grids
\citep{sandquist1998a,sandquist2000a,ricker2008a,ricker2012a,passy2012a,ohlmann2016a,ohlmann2016b,staff2016b,staff2016a,iaconi2016a}
or smoothed particle hydrodynamics
\citep[SPH,][]{passy2012a,nandez2014a,nandez2015a}. 
As an approximation to the giant structure, the core of the giant is usually
replaced by a point mass and the remaining envelope profile is mapped to the
grid without modifications
\citep[e.g.,][]{sandquist1998a,sandquist2000a,passy2012a,staff2016b,staff2016a,iaconi2016a}.
This procedure leads to mismatches in the hydrostatic equilibrium of the
envelope, resulting in spurious velocity fluctuations that have to be damped
away. The resulting giant profile has to be stable for several dynamical
timescales to ensure that modeling the spiral-in is not dominated by spurious
artifacts.  However, the procedures of approximating the giant profile, mapping
it to the hydrodynamical grid and subsequently relaxing it have not been analyzed
in detail so far.

In this paper, we present a novel method of approximating and reconstructing
giant profiles and show that they can be mapped to and relaxed in
multi-dimensional hydrodynamical simulations. This procedure leads to stable
giant envelopes that can reliably be used for hydrodynamical simulations of the
CE phase.  Similar to previous approaches, the core of the giant is replaced by
a point mass with a softened gravitational potential. Instead of a simple
mapping, however, the profile of the giant envelope is reconstructed by
integrating the hydrostatic equilibrium, taking into account the self-gravity of
the gas and the gravity of the point mass. This way, the reconstruction does
not depend on the grid resolution, and it can specifically control the
convective behaviour of the profile. Using a relaxation procedure that damps
out spurious velocities, models are obtained in hydrodynamical simulations that
are stable over several dynamical timescales. They are then used as initial
conditions for simulations of the CE phase, such as those presented in
\citet{ohlmann2016a,ohlmann2016b}.

The structure of the paper is as follows: we introduce the numerical methods in
Sec.~\ref{sec:methods}. Approximations to giant structures and reconstruction
methods are explained in Sec.~\ref{sec:stellarstructures}. Results from
hydrodynamical simulations are presented in Sec.~\ref{sec:hydrosimulations}.
After a discussion (Sec.~\ref{sec:discussion}), we conclude in
Sec.~\ref{sec:conclusions}.

\section{Numerical methods}
\label{sec:methods}

The hydrodynamical simulations presented in this paper are carried out using the
moving-mesh code \textsc{arepo} \citep{springel2010a}. It solves the Euler
equations with a finite volume approach on an unstructured, moving Voronoi mesh.
Moreover, the mesh can be adaptively refined on arbitrary criteria; here, we
employ this capability to ensure a similar cell mass on average. A tree method
solves for the self-gravity of the gas. The numerical method has recently been
updated to ensure second order convergence on arbitrary moving meshes
\citep{pakmor2016a}. As equation of state, either an ideal gas equation of state
(EOS) for a monoatomic gas with $\gamma=5/3$ is utilized or the OPAL EOS
\citep{rogers1996a,rogers2002a}, which incorporates the ionization state of the
medium and also radiation pressure. Radiative transport is not included
in our calculations.

The stellar evolution code \textsc{mesa} \citep{paxton2011a,paxton2013a,paxton2015a} is
used in version 6208 to create stellar models in various evolutionary stages.
We set the metallicity to $Z=0.02$ and use standard settings otherwise, i.e.,
the Reimers prescription with $\eta = 0.5$ for red giant (RG) winds
\citep{reimers1975a} and the Bl\"ocker prescription with $\eta = 0.1$ for
asymptotic giant branch (AGB) winds \citep{bloecker1995a}. These values are the
same as in \citet[see there for further information and
references]{paxton2013a}.

\subsection{Computational grids}
\label{sec:grids}

Because \textsc{arepo} solves the hydrodynamics equations on an unstructured
grid, we have the freedom to adapt the grid to the problem at hand. Different
types of initial grids are implemented:
\begin{itemize}
  \item Regular cubic grids.
  \item Nested cubic grids, where for each level of nesting 
    the inner half of the grid in each direction (i.e. an eighth of the total
    volume) is replaced by a grid with double the resolution and half the grid
    spacing of the original level.
  \item An adaptive cubic grid, where the grid is built in spherical shells
    tracking the mass distribution of the star to create approximately 
    equal-mass cells. Given the mass profile $m(r)$ of the star and the mass of
    cells $m_\mathrm{cell}$, the shells are constructed as follows:
    for each inner radius $r_1$, we compute an outer radius $r_2$
    and the grid spacing $a$ using conditions on the mass of the shell,
    \begin{equation}
      m(r_2) - m(r_1) = N m_\mathrm{cell},
      \label{eq:masscond}
    \end{equation}
    and on the volume of the shell,
    \begin{equation}
      \frac{4\pi}{3} \left( r_2^3 - r_1^3 \right) = N a^3,
      \label{eq:volcond}
    \end{equation}
    where $N$ is the number of cells in the shell. As an additional
    constraint, we impose that the width of the shell is a multiple of the grid
    spacing,
    \begin{equation}
      r_2 - r_1 = k a,
      \label{eq:radcond}
    \end{equation}
    where we usually choose $k=5$. These equations are solved subsequently for
    each shell to yield $r_2$, $a$, and $N$. In each shell, grid points are
    placed on a cubic grid with spacing $a$, where only the points with radii
    between $r_1$ and $r_2$ are considered.
  \item A HEALPix grid with spherical shells tracking the mass profile of the
    star, where the cells are distributed in each shell according to the HEALPix
    tesselation \citep{gorski2005a} similar to the distribution of SPH particles
    by \citet{pakmor2012b}. We deviate slightly from their scheme as we are not
    bound to strictly equal-mass cells compared to particles in SPH simulations.

    In each shell ranging from a given radius $r_1$ to $r_2$, we distribute
    $12 N_\mathrm{side}^2$ cells using the algorithm of \citet{gorski2005a},
    where $N_{\mathrm{side}}$ denotes the number of divisions of a base pixel in
    the HEALPix algorithm.
    Thereby, it is ensured that the cells are symmetric and have
    approximately a mass of $m_{\mathrm{cell},0}$. The cells have a radial
    diameter of $\Delta r_\mathrm{r} = r_2 - r_1$, whereas the lateral
    diameter can be computed by multiplying the midpoint $(r_1 + r_2) / 2$
    by the angular resolution $\theta$ of the shell \citep{gorski2005a},
    \begin{equation}
      \Delta r_\mathrm{l} = \frac{r_1 + r_2}{2} \theta =
      \frac{r_1 + r_2}{2} \sqrt{\frac{\pi}{3}}
      \frac{1}{N_\mathrm{side}}.
      \label{rq:horizontaldiameter}
    \end{equation}
    Let the symmetry factor $S$ be $S = \Delta r_\mathrm{r} / \Delta
    r_\mathrm{l}$, which is one for symmetric cells. By imposing $S=1$,
    $r_2$ can be computed for different values of $N_\mathrm{side}$. We then
    chose the radius for which the resulting cell mass,
    \begin{equation}
      m_\mathrm{cell} = \frac{m(r_2) - m(r_1)}{12 N_\mathrm{side}^2},
      \label{eq:cellmass}
    \end{equation}
    is closest to the given cell mass $m_{\mathrm{cell},0}$. To avoid
    large jumps in this quantity, it is limited by
    \begin{equation}
      \frac{1}{1.3} < \frac{m_\mathrm{cell}}{m_{\mathrm{cell},0}} < 1.3.
      \label{eq:massbound}
    \end{equation}
    If the mass limit is enforced for a shell (i.e., when condition
    (\ref{eq:massbound}) is violated), the number
    $N_\mathrm{side}$ is used for which $\log(S)$ is closest to zero (i.e., 
    the most symmetric cells are chosen).
    To resolve the steep gradients near the surface of the star, the mass limit
    is not enforced for radii larger than 95\% of the radius of the star.
    The positions of the cells on each sphere are computed according to
    the HEALPix algorithm that needs a coordinate system. This system is rotated
    randomly in space for each shell to avoid grid orientation effects.
\end{itemize}

For each grid, the density and pressure profiles were continued outside the
stellar surface at sufficiently low values to emulate the vacuum there in a
numerically feasible way. For the adaptive cubic and HEALPix grids, a
face-centered cubic grid was added in the outer part to sample the vacuum values
in the rest of the box. 
The box size is chosen to be four times as large as the radius of the
corresponding stellar model. Periodic boundary conditions are employed, because
no outflows are expected and the boundary is far enough away. Tests with larger
boxes did not show different results.

Because \textsc{arepo} uses an adaptive and moving mesh,
the choice of the initial grid does not significantly influence the simulations
(at least if it is not too coarse in the beginning); after a few time steps, the
configuration is similar for each initial grid (see
Sec.~\ref{sec:numericalparameters}).  If the mesh remains static, however,
mass-adaptive meshes are superior to regular meshes.  Except for three
simulations discussed in Sec.~\ref{sec:numericalparameters}, the HEALPix grid is
used for all simulations presented in this paper.

\subsection{Mapping stellar structures onto a grid}
\label{sec:mapping}

The initial conditions for hydrodynamical simulations are generated by mapping a
one-dimensional, spherically symmetric profile from the stellar evolution code
onto the grid used by the hydrodynamics code. In a finite volume method, as also
employed by \textsc{arepo}, the average values of the conserved quantities are
stored in each cell. Thus, we integrate over spherical shells when using the
HEALPix grid to compute the averages of density and internal energy from the
stellar evolution profile; these are then used as input values for the
hydrodynamics code. It is also possible to integrate over the pressure and
compute the internal energy from the EOS\@. This method may reproduce the
hydrostatic equilibrium slightly better, but we found the differences to be very
small ($\lesssim 1\%$ in most parts of the envelope), hence, we use the internal
energy from integrating over the shells.

For the other grids, it is more difficult to compute integrals over the cell
volumes because the geometry of the Voronoi mesh has to be known. The integral
can be approximated by taking the value of the spherically symmetric
function at the radius that is given by the center of mass of the cell. Since
the cubic grids are mostly regular (except at the interface between nested
levels in the nested grid or between adjacent spherical layers in the adaptive
mesh), this can also be approximated by the value computed at the
mesh-generating point of the cell. Thus, for all other grids besides the HEALPix
grid, the value of the density and internal energy are taken at the position of
the mesh-generating point of the corresponding cell. The differences in the
simulations between the grids, however, are small due to using an adaptive and
moving mesh (see Sec.~\ref{sec:numericalparameters}).

\subsection{Relaxation procedure}
\label{sec:relaxation}

Mapping spherically symmetric stellar models to multi-dimensional hydrodynamical
grids introduces discretization errors in the hydrostatic equilibrium. This is
primarily due to discretizing the pressure term differently from the
gravitational term in the numerical scheme: pressure enters through flux
computations in the finite volume scheme whereas gravity is computed point-wise
in the tree method. An additional source of error is the interpolation from the
high-resolution stellar profile to the coarser hydrodynamical grid.  Thus, errors
in the hydrostatic equilibrium are introduced leading to spurious velocities. 

The magnitude of the velocity fluctuations caused by resolution effects
can be estimated for a simple one-dimensional example of an isothermal
atmosphere with pressure scale height $H$. 
This analytical estimate assumes that the velocities are zero initially
and yields the amplitude of velocity fluctuations after one timestep of the
numerical scheme.
For the atmosphere, the
update of the momentum can be computed according to the finite volume scheme:
the discrete pressure is expanded into a Taylor series in the grid spacing
$\delta$, where the first order term cancels the gravitational source term. The
higher order terms depend on the numerical flux function, i.e., on the Riemann
solver and on the reconstruction.  For the \textsc{arepo} scheme, these terms
lead to spurious velocities that can be expressed in terms of the Mach
number $M$ for the first part of a time step as (for details of the computation,
see Appendix~\ref{sec:derivation})
\begin{equation}
  M = \frac{C_\mathrm{CFL}}{12\gamma} \left(
  \frac{\delta}{H} \right)^3 + O \left( \left( \frac{\delta}{H} \right)^4
  \right),
  \label{eq:machfluctuations}
\end{equation}
where $\gamma$ is the adiabatic index and $C_\mathrm{CFL}$ the
Courant-Friedrichs-Levy constant that appears in the stability constraint for
the time step $\Delta t = C_\mathrm{CFL} \delta / c_\mathrm{s}$.
Although this result is strictly only valid in the one-dimensional case, it
gives a lower limit for the resolution that is needed also in multi-dimensional
simulations to avoid velocity fluctuations due to the spatial discretization. 

Thus, to stabilize a hydrostatic atmosphere down to Mach numbers of 0.01
(0.001), one pressure scale height has to be resolved with at least 3.2 (6.8)
cells (assuming $\gamma = 5/3$ and $C_\mathrm{CFL} = 0.4$). This is of course
only a necessary but not a sufficient condition for reaching the
corresponding amplitude of Mach number fluctuations.
We ensure that all our initial models fulfill this criterion and
have sufficient resolution per pressure scale height.

Even if the resolution requirement is met, other sources of numerical error
(e.g., interpolation error, errors from the gravity solver) still introduce
spurious velocity fluctuations. Thus, an appropriate relaxation procedure
is necessary when mapping stellar models to hydrodynamical grids: the
velocity fluctuations have to be damped. This can be reached by adding a term in
the momentum equation proportional to
\begin{equation}
  \dot{\vec{v}} = - \frac{1}{\tau} \vec{v}.
  \label{eq:dampingpde}
\end{equation}
It damps the velocities on a timescale $\tau$.  The time discretization
is the same as for the rest of the scheme, and Eq.~(\ref{eq:dampingpde}) is
simply added as an additional source term.

The value of the damping timescale is essential for reaching stable models,
especially for dilute giant envelopes that encompass many timescales. The
following scheme, based on \citet{pakmor2012b} and \citet{rosswog2004a}, proved
to produce stable models: the physical time of the whole relaxation run is
chosen to be $10t_\mathrm{dyn}$ where the dynamical timescale $t_\mathrm{dyn}$
can either be the sound crossing time or the free-fall timescale (which are
usually of similar order); we use the sound crossing time of the stellar models.
For the value of the damping timescale, we use a small value $\tau_1 =
t_\mathrm{dyn}/10$ in the beginning that is subsequently increased to $\tau_2 =
t_\mathrm{dyn}$ (i.e., the damping is decreased) according to
\begin{equation}
  \tau(t) = \begin{cases}
    \tau_1, \quad & t < 2t_\mathrm{dyn} \\
    \tau_1 \left( \frac{\tau_2}{\tau_1} 
    \right)^{\frac{t-2t_\mathrm{dyn}}{3t_\mathrm{dyn}}}, \quad & 2t_\mathrm{dyn}
    < t < 5t_\mathrm{dyn} \\
    \infty, \quad & t > 5t_\mathrm{dyn} 
  \end{cases}.
  \label{eq:relaxationtime}
\end{equation}
The exponential form ensures that the temporal change of the damping timescale
is proportional to the damping timescale itself. This means that damping on
longer timescales (i.e., a larger value of $\tau$) is also applied for a longer
time. Taking $\tau_2$ on the order of the dynamical timescale is necessary to
suppress pulsations that typically occur on this timescale.  After
$5t_\mathrm{dyn}$, the damping is disabled and the system is evolved for another
$5t_\mathrm{dyn}$ to ensure that the stellar model is stable in
hydrodynamical simulations without damping.

\section{Stellar structures}
\label{sec:stellarstructures}

\subsection{Spatial structure of stars}

\begin{figure}[tbp]
  %\centering
  \includegraphics[width=\columnwidth]{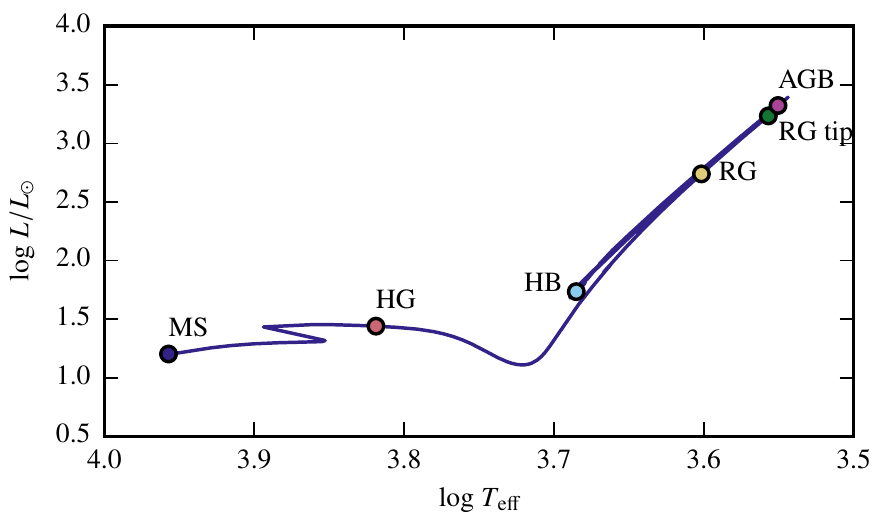}
  \caption{{Hertzsprung--Russell
          diagram for the evolution} of a $2 M_\odot$ star using \textsc{mesa}\@. Different
      evolutionary phases are marked for which the profiles are shown in
      Fig.~\ref{fig:2Mprofiles-relr}.}
  \label{fig:2Mhrd}
\end{figure}
\begin{figure}[tbp]
  %\centering
    \includegraphics[width=\columnwidth]{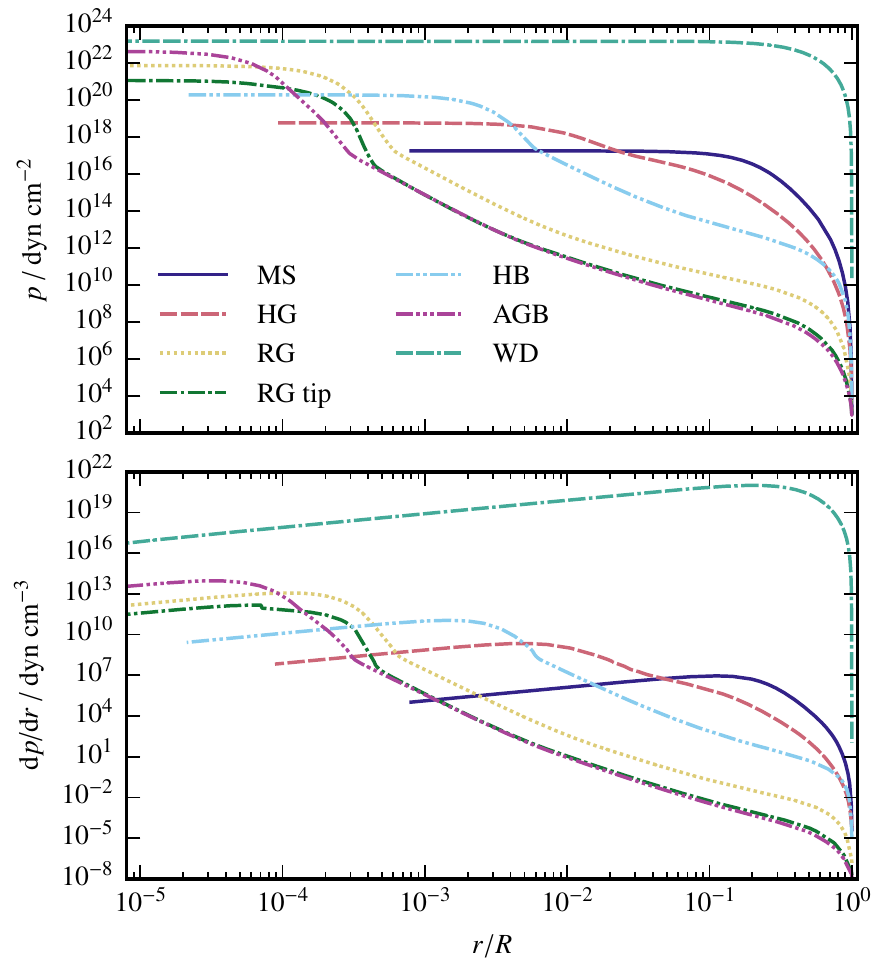}
  \caption{{Profiles
    of pressure $p$ and pressure derivative $\diff p/\diff r$} plotted over
      the relative radius $r/R$, where $R$ is the stellar radius of the
      corresponding profile. Shown are different evolutionary states for a ZAMS
      of $2 M_\odot$ and an isothermal WD of $0.6M_\odot$.}
  \label{fig:2Mprofiles-relr}
\end{figure}

In hydrodynamical simulations the Euler equations are discretized in space and
time; thus, it is important to analyze the spatial structure of the stars that
are to be mapped to the computational grid. As an example, we compare stellar
profiles at different evolutionary stages for a model with a zero-age main
sequence (ZAMS) mass of $2M_\odot$, which is evolved until the first thermal
pulse on the AGB using \textsc{mesa}\@. The evolution of the model is shown in a
Hertzsprung--Russell diagram in Fig.~\ref{fig:2Mhrd}, where different
evolutionary points are marked for which the pressure profiles are given in
Fig.~\ref{fig:2Mprofiles-relr}. For comparison, a profile of the interior of a cold, isothermal
$0.6M_\odot$ carbon-oxygen WD with $T=5\times10^5$~K is shown as well. This
model was set up by integrating the hydrostatic equilibrium equations with the
Helmholtz equation of state \citep{timmes2000a} starting from a central density
of $3.2\times 10^6$~g~cm$^{-3}$.

In Fig.~\ref{fig:2Mprofiles-relr}, the profiles of the pressure and its spatial
derivative are compared as functions of relative radius $r/R$, where $R$ is the
stellar radius of the corresponding model. 
The models show the usual behaviour: the core of the star contracts while its
envelope expands when it evolves away from the MS\@.  During core He burning on
the horizontal branch (HB), the core expands and the envelope contracts
slightly. On the AGB, the envelope expands even further than on the first red
giant branch (RGB) and the maximum pressure in the core rises.  The RGB and AGB
profiles are similar, especially in the outer parts.  The pressure profiles
(Fig.~\ref{fig:2Mprofiles-relr}, upper panel) show that the range in scales in
radius and pressure increases drastically during the evolution. In the relative
radius, it increases from roughly one to two decades for the MS model to five
decades for the AGB model.  The pressure derivative
(Fig.~\ref{fig:2Mprofiles-relr}, lower panel) shows a linear behaviour for $r\to
0$, which is expected from expanding the stellar structure equations about the
origin. To obtain stable models, the profile has to be resolved to the scale
where the pressure derivative decreases linearly; otherwise a cusp in the
pressure would be introduced at the origin, leading to spurious velocities.
Resolving these scales is computationally not feasible for more evolved stellar
structures with current numerical techniques. 

Thus, approximations are needed to model giant profiles in hydrodynamical
simulations -- as opposed to MS and WD models, which can be mapped to a
hydrodynamical simulation without approximations due to the smaller range in
scales.

\subsection{Approximations to giant profiles}

Giant profiles are approximated here by replacing the core of the giant by a
point mass that interacts only gravitationally with the envelope.
Gravitation-only particles have been implemented in \textsc{arepo} for modeling
dark matter \citep{springel2010a}. Their gravitational acceleration is given by
a spline function \citep[see also their Fig.~15]{springel2010a} 
\begin{equation}
  g_\mathrm{c}(r) = - G m_\mathrm{c} \frac{r}{h^3} 
  \begin{cases}
    -\frac{32}{3} + u^2 \left( \frac{192}{5} - 32 u \right), & 0 \leq u <
    \frac{1}{2}, \\
    \frac{1}{15 u^3} - \frac{64}{3} + 48 u \\ - \frac{192}{5}u^2 +
    \frac{32}{3} u^3, & \frac{1}{2} \leq u < 1, \\
    - \frac{1}{u^3},  & u \geq 1,
  \end{cases}
  \label{eq:spline}
\end{equation}
where $u = r/h$, $h$ is the softening length of the interaction, and
$m_\mathrm{c}$ is the mass of the particle representing the core. One advantage
of this functional form is that it reduces exactly to the Newtonian
acceleration outside the softening radius (i.e., for $r>h$) while still
avoiding the singularity at $r=0$. For radii smaller than $h$, gravity is
modified; thus, the stellar profile of the giant is cut out and has to be
replaced by a suitable function in this central region. Throughout this paper,
we choose $g_{\mathrm{c}}$ as given by Eq.~(\ref{eq:spline}), but the method
can also be used for other functional forms of $g_{\mathrm{c}}$.

To integrate the giant profiles including the gravitation of the point mass, the
stellar structure equations are modified,
\begin{equation}
  \begin{split}
    \frac{\diff m(r)}{\diff r} &= 4\pi \rho(r) r^2, \\
    \frac{\diff p(r)}{\diff r} &= - G \frac{m(r) \rho(r)}{r^2} - \rho(r) g_\mathrm{c}(r),
  \end{split}
  \label{eq:stellarstructurepressurenew}
\end{equation}
where $g_\mathrm{c}$ is the core acceleration given in Eq.~(\ref{eq:spline}),
$m$ denotes the integrated envelope mass (without the core mass), $\rho$ the
density of the gas, and $p$ the pressure. Thus, the self-gravity of the gas as
well as the gravity of the core is taken into account; the function $\rho(r)$
still has to be specified.

\begin{figure*}[tbp]
  \centering
  \includegraphics[width=\textwidth]{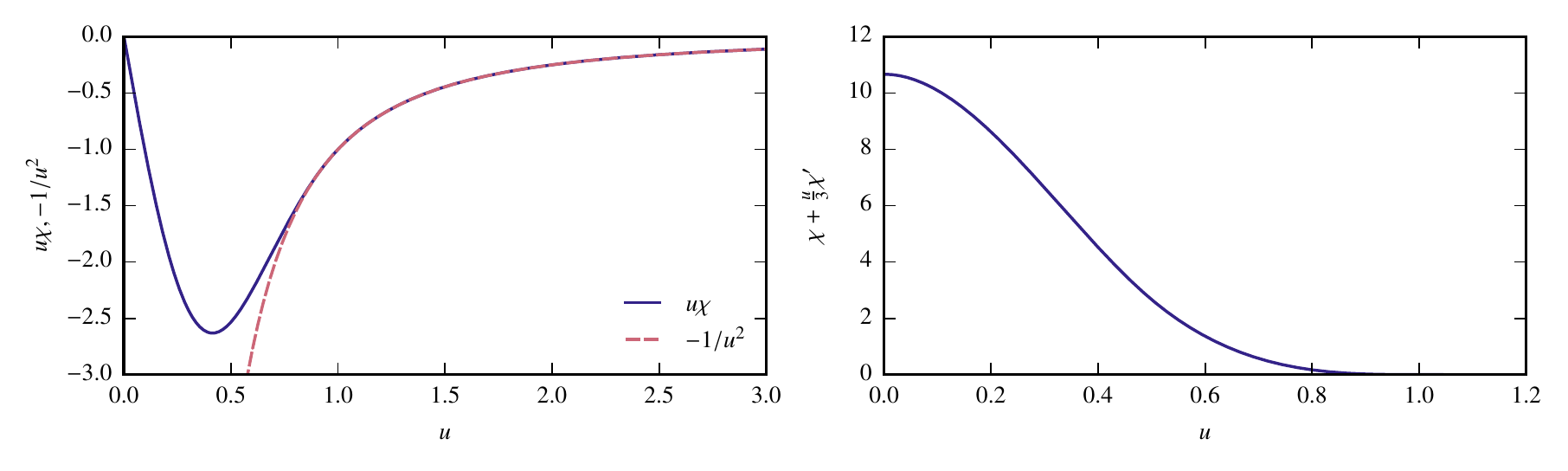}
  \caption{Approximate gravitational acceleration (left) and corresponding
    density for the modified Lane--Emden equation (right). The left panel shows
    $u\chi(u)$ ($\chi$ from Eq.~\ref{eq:spline}) compared to a Newtonian
    acceleration of a point mass ($-1/u^2$). Above $u=1$, both functions coincide by
    construction.  The right panel shows $\chi(u) + \frac{u}{3}\chi'(u)$, the
    dimensionless density distribution from Eq.~(\ref{eq:laneemdensplit2})
    associated with a softened gravitational acceleration as shown in the left
    panel.}
  \label{fig:compsoftening}
\end{figure*}

To create a stellar profile from the center of the star to the surface, the
profile is continued from the cut radius $h$ inwards to replace the original
profile in the central region.  The requirements for this continuation are that
it should be smooth, lead to a similar profile when mapped into the
hydrodynamics code and thus should be aware of the presence of the gravitational
force of the core.

To obtain continuations that fulfill these requirements, we introduce a modified
Lane--Emden equation (cf., e.g., \citealp{kippenhahn2012a} for a discussion of
the Lane--Emden equation) that can be derived by combining the modified stellar
structure equations~(\ref{eq:stellarstructurepressurenew}) to yield
\begin{equation}
  \frac{1}{r^2} \frac{\diff}{\diff r} \left( \frac{r^2}{4\pi G \rho}
  \frac{\diff p}{\diff r} + \frac{r^2 g_c(r)}{4\pi G} \right) + \rho = 0,
  \label{eq:modlaneemdendimensional}
\end{equation}
where $g_c$ is the gravitational force of the core given by
Eq.~(\ref{eq:spline}). One can now introduce a polytropic relation
$p=K\rho^{1+\frac{1}{n}}$ (with $K$ denoting the  polytropic constant, and $n$
the polytropic index), and
$\rho = \rho_0 \theta^{n}$ with the polytropic temperature $\theta$ and the
central gas density $\rho_0$, and a
scaled radius
\begin{equation}
  \xi=r/\alpha \quad \mathrm{with} \quad
  \alpha^2 = \frac{K(n+1)\rho_0^{\frac{1}{n} - 1}}{4 \pi G }.
  \label{eq:alpha}
\end{equation}
Inserting these relations into Eq.~(\ref{eq:modlaneemdendimensional}) yields the
well-known Lane--Emden equation, with an additional term that accounts
for the core,
\begin{equation}
  \frac{1}{\xi} \frac{\diff}{\diff \xi} \left( \xi^2
  \frac{\diff\theta}{\diff\xi} + \xi^2 \frac{g_c(\alpha\xi)}{4\pi G\rho_0 \alpha} \right)
  + \theta^n = 0.
  \label{eq:laneemdennondimensional1}
\end{equation}
By introducing a new function $\chi$,
\begin{equation}
  \chi \left( \frac{\xi}{\xi_h} \right) = \frac{h^3}{G m_c \alpha \xi}
  g_c(\alpha \xi),
  \label{eq:functionchi}
\end{equation}
where $\xi_h=h/\alpha$ is the non-dimensional radius at which the profile is
cut, Eq.~(\ref{eq:laneemdennondimensional1}) can be written as
\begin{equation}
  \frac{1}{\xi^2} \frac{\diff}{\diff\xi}\left( \xi^2 \frac{\diff
  \theta}{\diff\xi} \right) + \frac{\overline{\rho}}{\rho_0} 
  \left( \chi(\xi/\xi_h) + \frac{\xi}{3\xi_h} \chi'(\xi/\xi_h) \right)
  + \theta^n = 0,
  \label{eq:modlaneemden}
\end{equation}
where $\overline{\rho} = m_c / (4\pi h^3/3)$ can be interpreted as a mean
density of the core (i.e., the core mass spread out in a sphere of radius
$h$). The function $u\chi(u)$ can be seen as a dimensionless form
of the gravitational acceleration in Eq.~(\ref{eq:spline}) and is compared in
Fig.~\ref{fig:compsoftening} (left panel) to the original Newtonian acceleration
$1/u^2$. The interesting point is that the gravitational force of the core
appears as an additional density term that depends on the ratio of the mean
density of the core to the central density of the polytropic solution. This can
be seen even more clearly when converting Eq.~(\ref{eq:modlaneemden}) into two
first-order ODEs that can be used for integrating,
\begin{equation}
    \frac{\diff\theta}{\diff\xi} = - \frac{\eta}{\xi^2}, 
  \label{eq:laneemdensplit1}
\end{equation}
\begin{equation}
    \frac{\diff\eta}{\diff\xi}   = \xi^2 \left[ \theta^n + 
    \frac{\overline{\rho}}{\rho_0} \left( \chi(\xi/\xi_h) +
    \frac{\xi}{3\xi_h} \chi'(\xi/\xi_h) \right) \right],
  \label{eq:laneemdensplit2}
\end{equation}
where $\eta$ is the non-dimensional mass and where the initial conditions are
given by $\eta(0)=0$, $\theta(0) = 1$. The original Lane--Emden equation does
not feature the second term in the brackets of Eq.~(\ref{eq:laneemdensplit2}).
This term in the mass integration corresponds to an additional density source
caused by the core particle.  Thus, assuming a smoothed gravitational force for
the core as given in Eq.~(\ref{eq:spline}) is similar to smearing out the mass
of the core inside a sphere of radius $h$ with a density profile as shown in
Fig.~\ref{fig:compsoftening} (right panel).

The solution of this modified Lane--Emden equation now depends not only on the
polytropic index $n$, but also on the parameters $\overline{\rho}/\rho_0$,
$\xi_h$, and on the form of the function $\chi$. For $\chi$ given by
Eq.~(\ref{eq:functionchi}) the corresponding term in Eq.(\ref{eq:modlaneemden})
reduces to zero for $\xi>\xi_h$, i.e., in the outer part. Thus, the modified
Lane--Emden equation is reduced to the usual Lane--Emden equation in this
region.  The behaviour of the solution in the inner part ($\xi<\xi_h$) depends
on the ratio $\overline{\rho}/\rho_0$. For $\overline{\rho}\ll \rho_0$, the
first term in the brackets in Eq.~(\ref{eq:laneemdensplit2}) dominates and the
impact of the modification is small. The density resulting from smearing out the
point mass, $\overline{\rho}$, can be neglected compared to the central gas
density, $\rho_0$ and the solution is similar to the solution of the Lane--Emden
equation. In the opposite case, $\overline{\rho}\gg \rho_0$, the second term in
the brackets in Eq.~(\ref{eq:laneemdensplit2}) dominates and the contribution of
the gas density is negligible in the inner part for integrating the solution.
Hence, the gas mass in the interior part is negligible compared to the mass of
the core, and the solution differs from the normal Lane--Emden equation. This is
usually the case for the models presented in this paper.

\begin{figure*}[tbp]
  \centering
  \includegraphics[width=\textwidth]{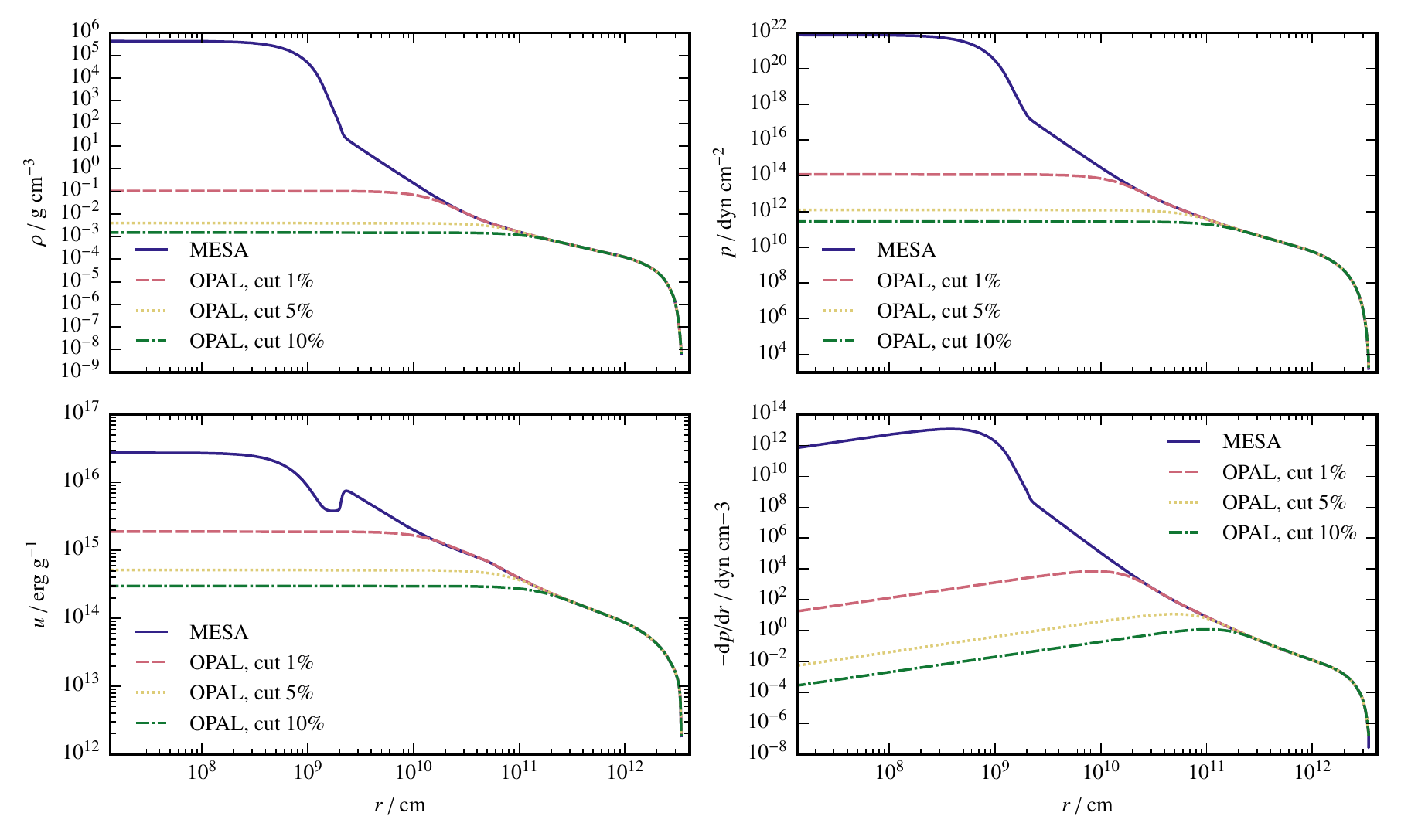}
  \caption{Comparison of density (upper left), pressure (upper right), internal
  energy (lower left) and derivative of pressure (lower right) for a $2 M_\odot$
  RG with a $\sim0.4M_\odot$ He core. Shown is the original profile from
  the \textsc{mesa} stellar evolution code as well as approximate profiles for cut radii
  of 1\%, 5\%, and 10\% of the total radius. The approximate profiles were
  computed using a polytropic index of $n=3$ for the interior part.}
  \label{fig:comprgdifferentsoftenings}
\end{figure*}

The solutions of the modified Lane--Emden equation show the same behaviour for
convective stability as the solutions of the original Lane--Emden equation: the
Schwarzschild criterion yields convective stability for
\begin{equation}
  n > \frac{1}{\gamma - 1}.
  \label{eq:schwarzschildlaneemden}
\end{equation}
For an adiabatic index $\gamma = 5/3$ corresponding to an ideal monoatomic gas,
this yields $n > 3/2$ for convectively stable stratifications.

To obtain a smooth transition to a stellar profile at a certain radius, the
solution of the modified Lane--Emden equation can be fitted to yield the
corresponding density and its derivative: the integration of
Eqs.~(\ref{eq:laneemdensplit1}) and (\ref{eq:laneemdensplit2}) yields the
function $\theta(\xi; n, h, \overline{\rho}; \alpha, \rho_0)$, where the
parameters $n, h$, and $\overline{\rho}$ are given. The density
\begin{equation}
  \rho(r; n, h, \overline{\rho}; \alpha, \rho_0) = \rho_0 
  \theta^n (\alpha r; n, h, \overline{\rho}; \alpha, \rho_0)
  \label{eq:modlaneemdendens}
\end{equation}
can be used to fit the parameters $\alpha$ and $\rho_0$ such that the density
and its derivative are matched at the transition radius. This has been
implemented using a non-linear root finder.

A few examples for such approximate stellar profiles are shown in
Fig.~\ref{fig:comprgdifferentsoftenings} for a $2M_\odot$ RG with a $\sim
0.4M_\odot$ He core (the RG model in Fig.~\ref{fig:2Mhrd}) for different core
cutoff radii.  In this example, a polytropic index of $n=3$ is used for the
modified Lane--Emden solution in the interior part that was fitted to the
profile. For the integration of the hydrostatic equilibrium of the profile, the
OPAL equation of state \citep{rogers1996a,rogers2002a} was used that is
identical to the equation of state used by \textsc{mesa} in this $\rho$-$T$ regime. The
plots show that the stellar profile is reproduced very well in the outer part
at radii larger than the cut radius for the important thermodynamic quantities.
The connection to the inner part of the modified Lane--Emden solutions is
smooth, as expected.  Moreover, the derivative of the pressure
(Fig.~\ref{fig:comprgdifferentsoftenings}, lower right) shows linear behaviour
for $r\to0$, which is important for ensuring hydrostatic equilibrium at the
center of the star.

\subsection{Reconstructing stellar profiles}
\label{sec:reconstruction}

To use a stellar profile from a stellar evolution code in a multi-dimensional
hydrodynamics code, the stellar structure equations have to be integrated when
employing a different equation of state or approximate profiles such as
described in the previous section. Especially for a different equation of
state, some of the thermodynamic quantities will deviate. In this case,
different reconstruction methods can be utilized to recover specific properties of
the models.

Usually, the density is taken from the stellar profile for
integrating the stellar structure equations: given $\rho(r)$ as function of the
radius, one can easily integrate 
\begin{equation}
  \begin{split}
    \frac{\diff m(r)}{\diff r} &= 4\pi r^2 \rho(r), \\
    \frac{\diff p(r)}{\diff r} &= - \frac{G m(r) \rho(r)}{r^2}.
  \end{split}
  \label{eq:integrategivenrho}
\end{equation}
This ensures that density, mass, and pressure profiles equal the values from
the stellar evolution code. However, the values of the internal energy, entropy,
sound speed, temperature gradients and other quantities may differ compared to
the stellar evolution code. Thus, regions of convective stability or instability
are not necessarily recovered correctly.

To conserve the convective stability or instability of the profile in different
regions, the difference $\nabla - \nabla_\mathrm{ad}$ between the temperature
gradient and the adiabatic gradient can be integrated in an alternative
way.\footnote{This idea was originally proposed by \citet{edelmann2016a},
although with fixed gravity.} To this end, a third equation
for the temperature has to be added that is integrated along with the other
equations,
\begin{equation}
  \begin{split}
    \frac{\diff m(r)}{\diff r} &= 4\pi r^2 \rho(p(r), T(r)), \\
    \frac{\diff p(r)}{\diff r} &= - \frac{G m(r) \rho(p(r), T(r))}{r^2}, \\
    \frac{\diff T(r)}{\diff r} &= \frac{\diff p(r)}{\diff r} \frac{T(r)}{p(r)} 
    \nabla (r),
  \end{split}
  \label{eq:integrategivennabla}
\end{equation}
where the temperature gradient is
\begin{equation}
  \nabla(r) \equiv \frac{\diff \ln T}{\diff \ln p}(r) = (\nabla - \nabla_\mathrm{ad})(r) + \nabla_\mathrm{ad}(p(r), T(r)),
  \label{eq:nabla}
\end{equation}
and $\nabla_\mathrm{ad}$ is the adiabatic gradient given by the equation of
state. This method ensures that the difference $\nabla - \nabla_\mathrm{ad}$
remains the same with respect to the corresponding equation of state. Hence, the
criterion for convective stability will be fulfilled in the same regions of the
stellar profile. This method, however, may change the mass profile of the star
and thus also its radius and total mass. Hence, the quantities that one wants to
preserve when mapping to a hydrodynamics code decide on the method to use.

\begin{figure*}[tbp]
  \centering
  \includegraphics[width=\textwidth]{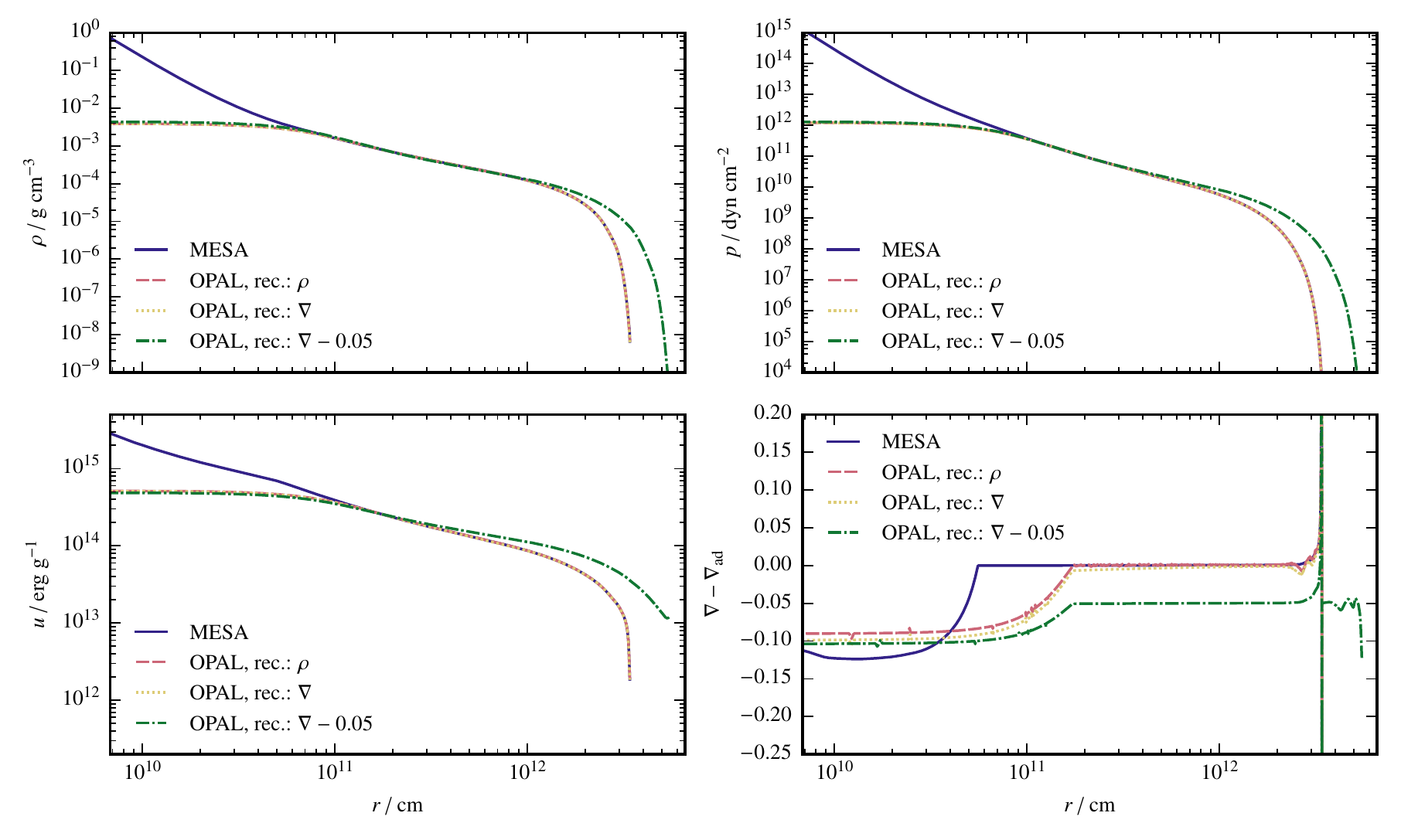}
  \caption{Comparison of different integration methods for a $2M_\odot$ RG using
  the OPAL equation of state. The profile was cut at 5\% of the stellar
    radius. 
  Shown are the density $\rho$ (upper left), the pressure $p$ (upper right), the
  internal energy $u$ (lower left), and the difference between the gradient
  $\nabla$ and the adiabatic gradient $\nabla_\mathrm{ad}$ (lower right).}
\label{fig:comprgstable}
\end{figure*}

\begin{figure*}[tbp]
  \centering
  \includegraphics[width=\textwidth]{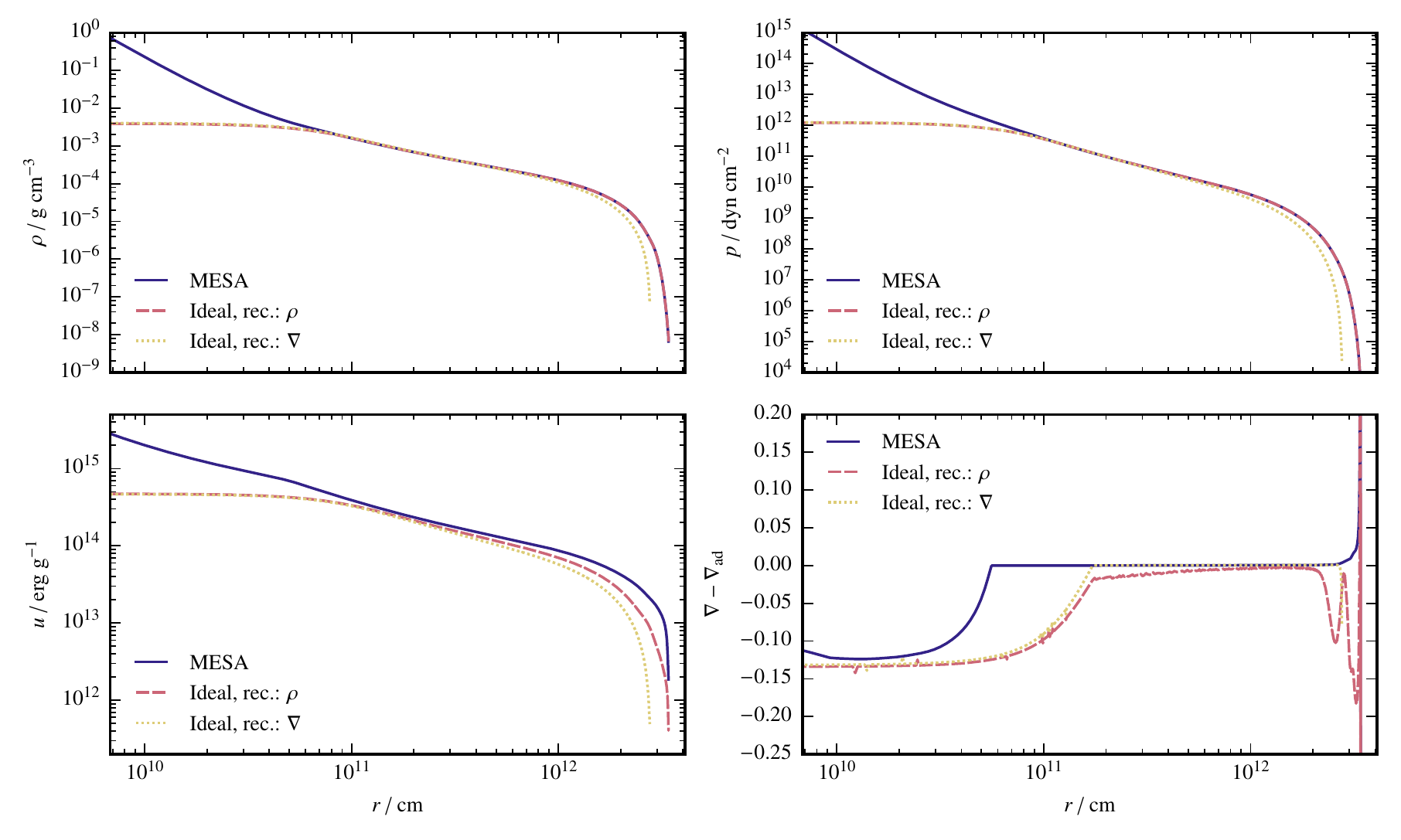}
  \caption{Comparison of different integration methods for a $2M_\odot$ RG using
  an ideal equation of state. The profile was cut at 5\% of the stellar radius.
  The panels show the same quantities as in Fig.~\ref{fig:comprgstable}.}
\label{fig:comprgideal}
\end{figure*}

When employing the same equation of state in the stellar evolution code and in the
hydrodynamics code, both integration methods yield the same result. This is
demonstrated in Fig.~\ref{fig:comprgstable} for a $2M_\odot$ RG that was cut at
5\% of the stellar radius. The figure shows the density $\rho$,
the pressure $p$, the internal energy $u$, and the difference between the
gradient $\nabla$ and the adiabatic gradient $\nabla_\mathrm{ad}$ for the
profile from \textsc{mesa} and reconstructed profiles using $\rho$ or $\nabla -
\nabla_\mathrm{ad}$ as given quantities. As one can see, both methods
yield almost the same result.

However, using an ideal gas EOS for the reconstruction yields different
profiles, as shown for the $2M_\odot$ RG in Fig.~\ref{fig:comprgideal}. The
deviations in the resulting stellar profiles is larger for extended giant
profiles because the difference between the ideal gas equation of state and the
OPAL equation of state is larger for lower densities and temperatures, where
recombination plays an important role. When reconstructing using the density
(red dashed line in Fig.~\ref{fig:comprgideal}), the profiles of density and
pressure are recovered accurately, while the internal energy is smaller in the
outer parts of the envelope. Moreover, $\nabla-\nabla_\mathrm{ad} < 0$ for most
parts of the envelope which means that the envelope is convectively stable when
using this profile for mapping to the hydrodynamics code in contradiction to the
original profile. The largest deviation can be seen in the outer part, where the
two dips in the profile of $\nabla-\nabla_\mathrm{ad}$ correspond to the
ionization zones of He and H. This model can also be reconstructed using
$\nabla-\nabla_\mathrm{ad}$ (Fig.~\ref{fig:comprgideal}, yellow dotted line),
reproducing the convectively unstable behaviour of the envelope for radii larger
than the cut radius. However, the profile constructed in this way is smaller in
radius by 18\% and in gas mass by 27\% (without core) because of the steeper
adiabatic gradient.

Apart from reproducing stellar models for a different equation of state, the
reconstruction method for $\nabla - \nabla_\mathrm{ad}$ also allows the
generation of stellar profiles for use in the hydrodynamics code that can
artificially be made convectively unstable or stable by adding or subtracting a
constant in Eq.~(\ref{eq:nabla}). The stratification of a convective envelope,
e.g., is nearly adiabatic. Changing $\nabla$ by a small amount (0.01-0.05, i.e.,
about 2.5\% to 12.5\% of the ideal gas adiabatic gradient of 0.4) leads to a
convectively stable stratification. Since the temperature gradient is now
shallower than in the original profile, the envelope will be larger in radius
and mass. 

As an example, this reconstruction method is applied to the same $2M_\odot$ RG
model in Fig.~\ref{fig:comprgstable} (yellow dotted line), with $\nabla$ being
decreased by 0.05.  Starting the integration from the center, this results in a
model with a convectively stable envelope because $\nabla <
\nabla_{\mathrm{ad}}$ there. However, the mass distribution deviates
from the original profile: the model is 61\% larger in radius and 88\% larger in
mass (without core mass). The spike visible in $\nabla - \nabla_{\mathrm{ad}}$
is located at the
surface of the original model and due to the gradient indicating
super-adiabicity in the very surface layers. This, however, affects only a small
part of the profile that cannot be resolved in the hydrodynamical simulations and
can thus be neglected.  Although the mass profile changes using this
reconstruction method, these models can be used for testing the numerical
stability in the hydrodynamical simulation (because convective motions should be
absent) while still having similar density and pressure profiles.

\section{Stable models in hydrodynamical simulations}
\label{sec:hydrosimulations}

\subsection{Stability criteria}

After creating the initial radial profiles -- either by taking the stellar
evolution data directly or by applying approximations -- they are mapped to a
three-dimensional grid to conduct hydrodynamical simulations. Discretization errors
resulting from the mapping or from the numerical scheme lead to velocity
fluctuations that may in turn lead to an unstable representation of the stellar
model. In this case, instabilities may emerge with large, potentially also
non-symmetric, spurious velocities resulting in expanded profiles or pulsating
envelopes.

The objective is thus to arrive at stable representations of stellar models
in hydrodynamical simulations. It is, however, difficult to define stringent
criteria on stability since inherent, convective motions are present in
convectively unstable envelopes, and hence the hydrostatic equilibrium cannot be
fulfilled exactly. For Mach numbers that are not too large ($\lesssim 0.1$), a
stationary state should result with only small deviations from the hydrostatic
equilibrium.

We require that after several dynamical time scales the
following approximate conditions are met in the hydrodynamical simulations:
\begin{enumerate}
  \item The deviations from the initial pressure and density profiles
    should be small.
  \item Both sides of the hydrostatic equilibrium equation,
     $\nabla p = - \rho \nabla \Phi$, should compensate each other.
  \item For convectively stable profiles, the Mach numbers in the model
    should be small compared to the Mach numbers expected in the problem.
  \item For convectively unstable models, a stationary state should be reached.
  \item To exclude the presence of pulsations, the potential energy of the
    system should be constant.
\end{enumerate}
To allow the hydrodynamics code to relax the initial profile to a stable
representation, sufficient numerical resolution is required and a damping scheme
is employed (see Sec.~\ref{sec:relaxation}).

\subsection{RG models}

The relaxation procedure as outlined in Sec.~\ref{sec:relaxation} is now applied
to RG structures to show that stable representations can be obtained
reliably. Since the space of stellar model parameters as well as numerical
parameters is huge, only a subset can be tested. To this end, a suite of 14
simulations was run for an initial model of a $2M_\odot$ RG in the middle of the
first giant branch (the same as in Fig.~\ref{fig:2Mhrd}). A summary of the
simulations is presented in Tab.~\ref{tab:relaxationruns}. In the following, we
discuss how well the original star is reproduced in terms of the mechanical
structure (i.e., density and pressure distributions) but also regarding
convectively stable or unstable layers.

\begin{table*}
  %\scriptsize
  %\footnotesize
  %\small
  \centering
  \caption{\textbf{Overview of RG Relaxation Runs.} The table shows data for 14
different simulations for the initial $2M_\odot$ RG model. The columns display a
model letter, 
the equation of state, the reconstruction method
(cf.~Sec.~\ref{sec:reconstruction}), the grid type (HP -- HEALPix, CU -- cubic,
CA -- cubic adaptive, CN -- cubic nested; cf.~Sec.~\ref{sec:grids}), the
core cutoff radius $r_\mathrm{c}$ as a fraction of the initial model radius, the
number of cells $N$, the radius and gas mass of the reconstructed profile in
solar units, the mean Mach number at the end of the relaxation run and the mean
relative difference in the hydrostatic equilibrium. The mean values are computed
using the cell mass as weight. The core mass in all simulations is
$0.38M_\odot$.}
\label{tab:relaxationruns}
  \begin{tabular}{llccS[table-format=1.2]S[table-format=1e1]S[table-format=2]S[table-format=1.1,table-auto-round]S[table-format=1.1e-1,table-auto-round]S[table-format=1.1e-1,table-auto-round]}
  %\toprule
  \hline\hline
                     Model &   EOS &    Recon. &           Grid & $r_c/R$ &
  {$N$} & {$\frac{R}{R_\odot}$} & {$\frac{M_\mathrm{env}}{M_\odot}$} & {$\overline{\mathcal{M}}$} &
  {$\overline{\frac{|\rho\vec{g} - \nabla p|}{\max(|\rho\vec{g}|,|\nabla p|)}}$} \\
  %\midrule
  \hline
A &  OPAL &            $\rho$ & HP &    0.05 &  2e6 & 49 & 1.61 & 1.18e-1 & 2.74e-2 \\  %&   OdHPr05n2e6 
B &  OPAL &            $\rho$ & HP &    0.05 &  5e5 & 49 & 1.61 & 1.16e-1 & 2.71e-2 \\  %&   OdHPr05n5e5 
C &  OPAL &            $\rho$ & HP &    0.05 &  1e5 & 49 & 1.61 & 7.50e-2 & 2.60e-2 \\  %&   OdHPr05n1e5 
D &  OPAL &            $\rho$ & HP &    0.02 &  5e5 & 49 & 1.61 & 1.33e-1 & 2.98e-2 \\  %&   OdHPr02n5e5 
E &  OPAL &            $\rho$ & HP &    0.10 &  5e5 & 49 & 1.60 & 8.60e-2 & 2.30e-2 \\  %&   OdHPr10n5e5 
F &  OPAL &   $\nabla - 0.02$ & HP &    0.02 &  5e5 & 62 & 2.78 & 1.87e-2 & 1.37e-2 \\  %& Og02HPr02n5e5 
G &  OPAL &   $\nabla - 0.02$ & HP &    0.05 &  5e5 & 59 & 2.07 & 9.56e-3 & 1.29e-2 \\  %& Og02HPr05n5e5 
H &  OPAL &   $\nabla - 0.02$ & HP &    0.05 &  2e6 & 59 & 2.07 & 9.14e-3 & 8.15e-3 \\  %& Og02HPr05n2e6 
I &  OPAL &   $\nabla - 0.05$ & HP &    0.05 &  5e5 & 79 & 3.02 & 5.32e-3 & 1.17e-2 \\  %& Og05HPr05n5e5 
J & Ideal &            $\rho$ & HP &    0.05 &  5e5 & 49 & 1.61 & 6.04e-3 & 1.28e-2 \\  %&   IdHPr05n5e5 
K & Ideal &          $\nabla$ & HP &    0.05 &  5e5 & 40 & 1.18 & 4.54e-2 & 1.55e-2 \\  %& Ig00HPr05n5e5 
L &  OPAL &            $\rho$ & CU &    0.05 &  5e5 & 49 & 1.61 & 1.16e-1 & 2.63e-2 \\  %&   OdCUr05n5e5 
M &  OPAL &            $\rho$ & CA &    0.05 &  5e5 & 49 & 1.61 & 1.12e-1 & 2.62e-2 \\  %&   OdCAr05n5e5 
N &  OPAL &            $\rho$ & CN &    0.05 &  5e5 & 49 & 1.61 & 1.09e-1 & 2.66e-2 \\  %&   OdCNr05n5e5 
  %\bottomrule
  \hline
  \end{tabular}
\end{table*}

% define model namees
%\newcommand{\modelA}[0]{OdHPr05n2e6~[A]}
%\newcommand{\modelB}[0]{OdHPr05n5e5~[B]}
%\newcommand{\modelC}[0]{OdHPr05n1e5~[C]}
%\newcommand{\modelD}[0]{OdHPr02n5e5~[D]}
%\newcommand{\modelE}[0]{OdHPr10n5e5~[E]}
%\newcommand{\modelF}[0]{Og02HPr02n5e5~[F]}
%\newcommand{\modelG}[0]{Og02HPr05n5e5~[G]}
%\newcommand{\modelH}[0]{Og02HPr05n2e6~[H]}
%\newcommand{\modelI}[0]{Og05HPr05n5e5~[I]}
%\newcommand{\modelJ}[0]{IdHPr05n5e5~[J]}
%\newcommand{\modelK}[0]{Ig00HPr05n5e5~[K]}
%\newcommand{\modelL}[0]{OdCUr05n5e5~[L]}
%\newcommand{\modelM}[0]{OdCAr05n5e5~[M]}
%\newcommand{\modelN}[0]{OdCNr05n5e5~[N]}
\newcommand{\modelA}[0]{model~A}
\newcommand{\modelB}[0]{model~B}
\newcommand{\modelC}[0]{model~C}
\newcommand{\modelD}[0]{model~D}
\newcommand{\modelE}[0]{model~E}
\newcommand{\modelF}[0]{model~F}
\newcommand{\modelG}[0]{model~G}
\newcommand{\modelH}[0]{model~H}
\newcommand{\modelI}[0]{model~I}
\newcommand{\modelJ}[0]{model~J}
\newcommand{\modelK}[0]{model~K}
\newcommand{\modelL}[0]{model~L}
\newcommand{\modelM}[0]{model~M}
\newcommand{\modelN}[0]{model~N}
\newcommand{\ModelA}[0]{Model~A}
\newcommand{\ModelB}[0]{Model~B}
\newcommand{\ModelC}[0]{Model~C}
\newcommand{\ModelD}[0]{Model~D}
\newcommand{\ModelE}[0]{Model~E}
\newcommand{\ModelF}[0]{Model~F}
\newcommand{\ModelG}[0]{Model~G}
\newcommand{\ModelH}[0]{Model~H}
\newcommand{\ModelI}[0]{Model~I}
\newcommand{\ModelJ}[0]{Model~J}
\newcommand{\ModelK}[0]{Model~K}
\newcommand{\ModelL}[0]{Model~L}
\newcommand{\ModelM}[0]{Model~M}
\newcommand{\ModelN}[0]{Model~N}

\begin{figure}[tbp]
  %\centering
  \includegraphics[width=\columnwidth]{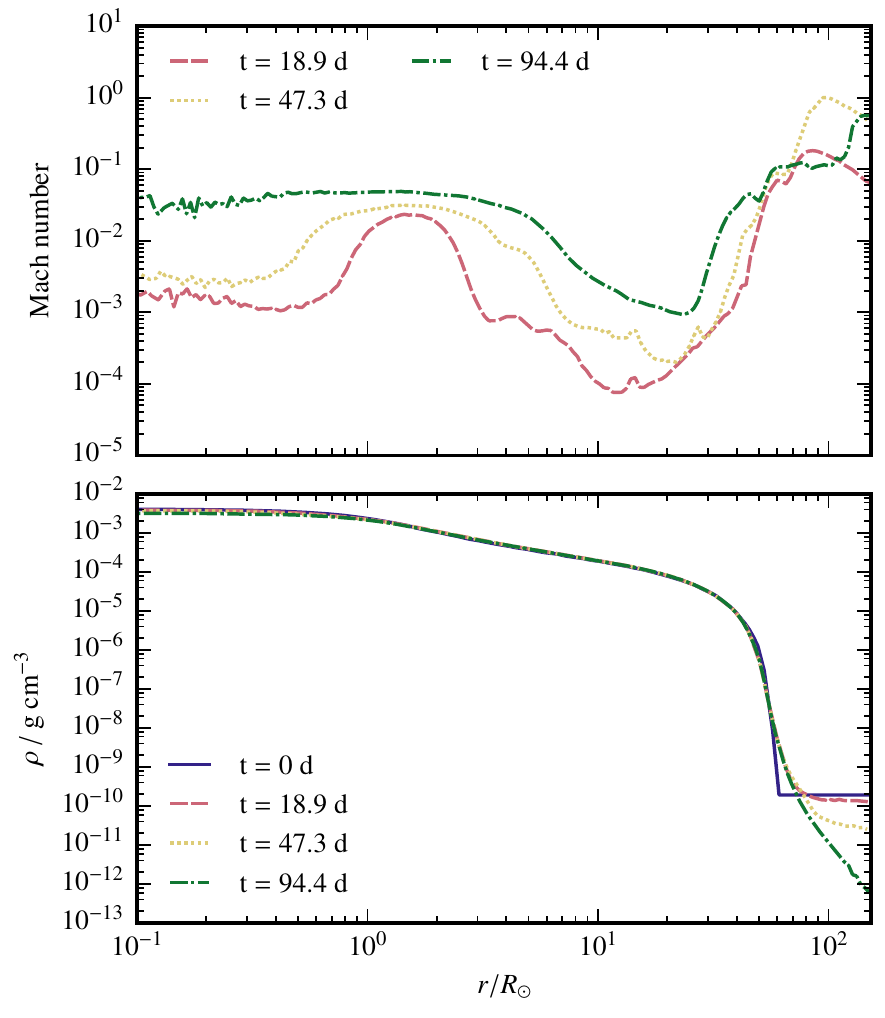}
  \caption{{Mach number and density profile for \modelH.}
  The top panel show the Mach number at different times during the relaxation
run, the bottom panel shows the density at the same times. Both quantities were
radially binned and averaged in shells. This model should be convectively stable.}
\label{fig:profileMrhoH}
\end{figure}

\begin{figure}[tbp]
  %\centering
  \includegraphics[width=\columnwidth]{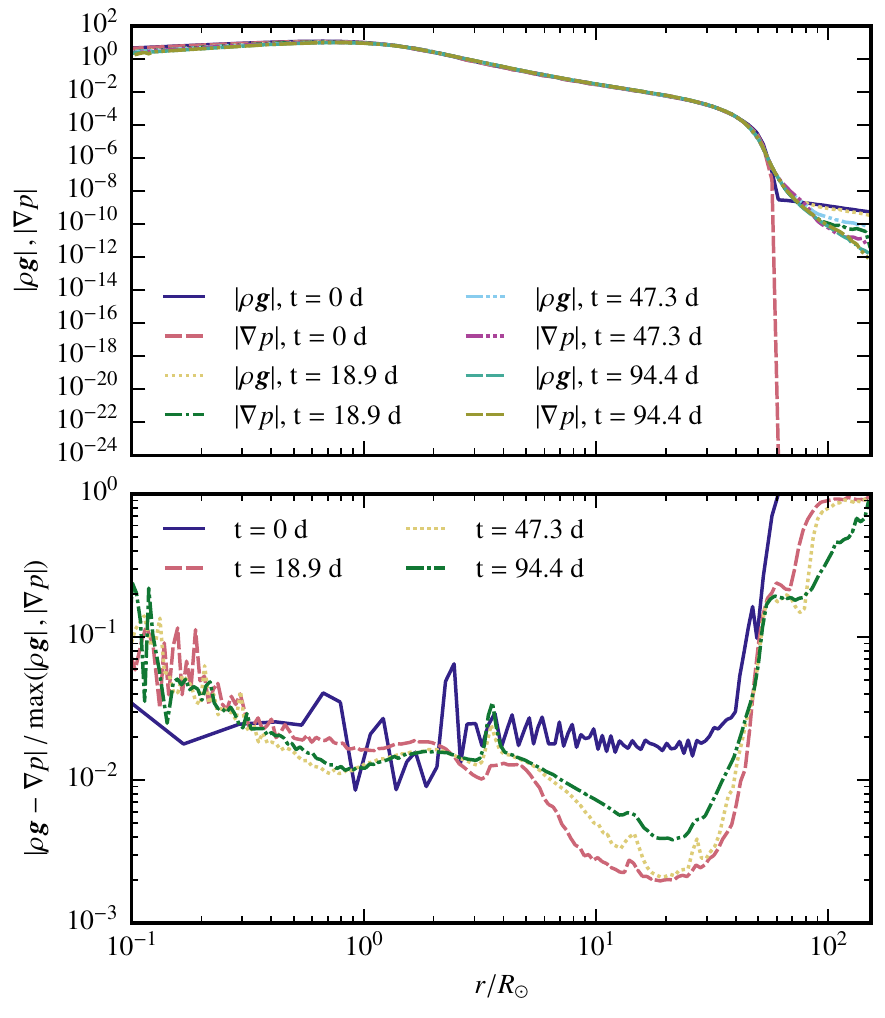}
  \caption{{Hydrostatic equilibrium for \modelH.} The top panel shows the two
  sides of the hydrostatic equilibrium, $\left|\nabla p\right|$ and $\rho
  \left|\vec{g}\right|$ at different times during the relaxation run. The bottom
  panel shows the relative error in the hydrostatic equilibrium. All quantities
  were radially binned and averaged in shells. This model should be convectively
  stable.}
\label{fig:profileHSEH}
\end{figure}

\begin{figure}[tbp]
  %\centering
  \includegraphics[width=\columnwidth]{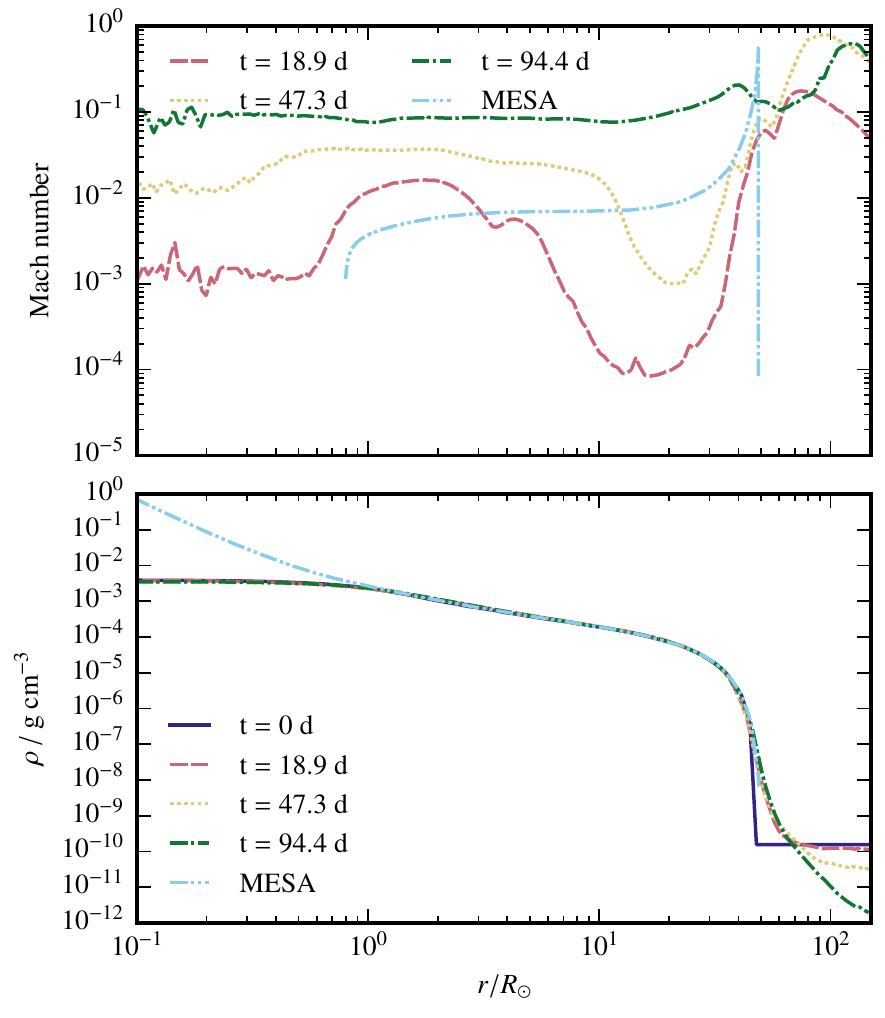}
  \caption{{Mach number and density profile for \modelA.}
  The top panel show the Mach number at different times during the relaxation
run, the bottom panel shows the density at the same times. Both quantities were
radially binned and averaged in shells. Moreover, the profile from the \textsc{mesa}
model is plotted. This model should be convectively unstable.}
\label{fig:profileMrhoA}
\end{figure}

\begin{figure*}[tbp]
  \centering
    \includegraphics[width=0.66\textwidth]{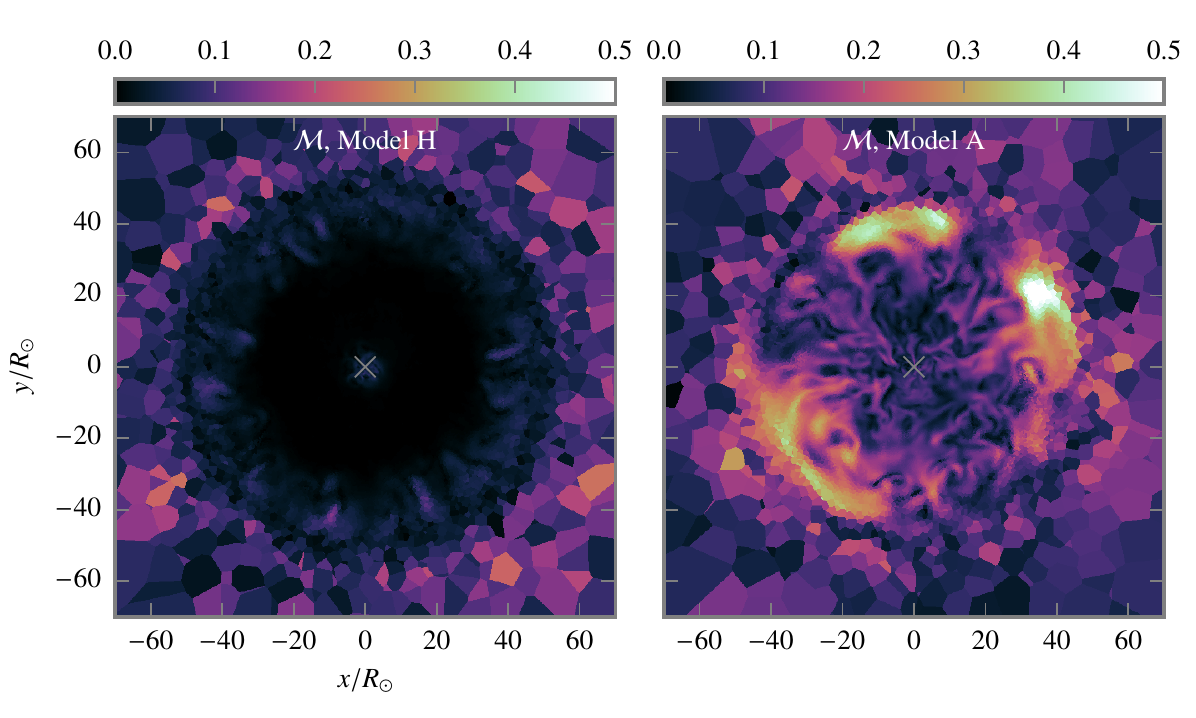}
  \caption{{Comparison of Mach number distributions for
      \modelH{} (left panel) and \modelA{} (right panel).} \ModelH{} should be
      convectively stable, \modelA{} convectively unstable. The low-density
    region outside the stars (large cells) shows spurious Mach number
    fluctuations.}
\label{fig:comparisonmachnumber}
\end{figure*}

\subsubsection{Convectively Stable Atmospheres}

The envelope of the $2M_\odot$ RG is convectively unstable and should thus show
convective motions. The magnitude of these velocities can be quite substantial:
the Mach number according to mixing-length theory as computed by \textsc{mesa} ranges up
to 0.5 in the outermost layers. When setting up such an initial model, it is
thus difficult to judge whether velocities are due to convective motions or
whether they
are numerical artifacts caused by the numerical scheme. Hence, the magnitude of
spurious velocities introduced by the numerical method is best studied for
convectively stable stars. We created a set of four artificial, convectively
stable models (\modelF, \modelG, \modelH, \modelI; see
Tab.~\ref{tab:relaxationruns}) by reconstructing the $2M_\odot$ RG model using a
temperature gradient shifted by a small amount for convective stability and
using the OPAL EOS, the same as in \textsc{mesa} (in this $\rho$-$T$ regime). This
way, models are created for which the density and pressure structure resembles
the initial stellar profile (see Fig.~\ref{fig:comprgstable}). The shift of 0.02
(0.05) corresponds to 5\% (12.5\%) of the adiabatic gradient for a value of 0.4
in the deeper interior.  The flatter temperature gradient, however, leads to
more extended profiles with more mass (Tab.~\ref{tab:relaxationruns} and
Fig.~\ref{fig:comprgstable}).

For the convectively stable \modelH{}, the Mach number and density
distributions are shown in Fig.~\ref{fig:profileMrhoH}, and the hydrostatic
equilibrium in Fig.~\ref{fig:profileHSEH}. As required for a stable model,
the density profile follows the initial profile closely
(Fig.~\ref{fig:profileMrhoH}, lower panel). In the central part, it
decreases by about 10\%, and at the surface, the very steep initial gradient
cannot be fully resolved, hence, a slight expansion occurs.
After the damping is switched off at $5t_{\mathrm{dyn}} =
\SI{47.3}{d}$,
the profile does not expand further. 
The Mach numbers are zero initially, increase during the relaxation run, and
level off at the end of the run (Fig.~\ref{fig:profileMrhoH}, upper panel). 
The model stays very subsonic with Mach numbers below 0.05 in
the inner part and less than 0.01 in most parts of the envelope. Near the
surface, the Mach numbers increase again up to 0.1. In the background regions
outside of the star, large spurious fluctuations occur. Because densities are
low in these regions, however, the fluctuations are unimportant for the
evolution of the stellar model (this also occurs for most of the other
simulations). The mean Mach number over the whole grid is
$0.009$. The Mach numbers
during the damping phase (at $\SI{18.9}{d}$) are largest near the surface and at
the radius where the core is connected to the envelope, i.e., at the softening
length of its gravitational interaction. From these regions, the fluctuations
disperse slowly into other regions of the envelope (compare red and yellow lines
in Fig.~\ref{fig:profileMrhoH}, upper panel; five dynamical timescales lie in
between those profiles).
An important point here is that the Mach numbers introduced during a
relaxation run in a convectively stable model (about 0.01) are much smaller
than those encountered in convectively unstable models (e.g., \modelA{}). Hence, the
convective motions seen in those models should not be affected by the imperfect
stabilization of the hydrostatic equilibrium.

The hydrostatic equilibrium for \modelH{} is shown in
Fig.~\ref{fig:profileHSEH}.
Both terms that should cancel in hydrostatic equilibrium are indeed very
similar and constant over time (Fig.~\ref{fig:profileHSEH}, upper
panel); small changes can be seen in the center (correlated to the decrease in
density there) and  near the surface due to the expansion. The error in the
hydrostatic equilibrium is shown in Fig.~\ref{fig:profileHSEH} (lower panel).
Initially, it is roughly 2\% throughout the envelope, but it decreases
during the relaxation run as the model converges to a hydrostatic equilibrium
that is stable in the hydrodynamical simulation. The decrease is largest between
$3R_\odot$ and $40R_\odot$ where most of the mass of the envelope is located.
The mean error of the hydrostatic equilibrium over the whole grid computed for
this model is $\num{8e-3}$ and hence, the final model fulfills the hydrostatic
equilibrium to better than a per cent on average.

The density profile is constant in time after disabling the damping and the Mach
number is small. This indicates that no pulsation is present, which
is confirmed by the potential energy showing very small fluctuations $<1\%$
after the initial expansion. Pulsations would cause regular changes in the
potential energy with a period of the dynamical timescale (about 9\,d for this
model). The initial expansion causes a small decrease of the potential energy by
2\% to 3\% (for all models). After this, the relative difference stays below 1\%
for all models in Tab.~\ref{tab:relaxationruns}. Moreover, the error in the
total energy is roughly 1\% for all models except \modelC{} (about 3\%)
due to its lower resolution. Thus, none of the models show significant
pulsations.

Combining all these results, we conclude that we are able to generate a model that
stays in hydrostatic equilibrium for a convectively stable profile. For
\modelG{} with a lower resolution, the results are very similar, only the
mean error in the hydrostatic equilibrium, $\overline{|\rho\vec{g} -
\nabla p|/\max(|\rho\vec{g}|,|\nabla p|)}$, is slightly larger with 1.3\%. For the
model with a smaller core cutoff radius, \modelF{}, the error in the hydrostatic
equilibrium is similar (1.4\%), but the resulting mean Mach number is larger:
0.02 instead of 0.01. Since the generation of the spurious velocities starts at
the boundary of the core, this indicates that higher resolutions are needed to
stabilize models with smaller core cutoff radii. The model that is farther away
from the threshold to convective instability, \modelI{}, shows a similar
hydrostatic equilibrium error and even lower Mach numbers with a mean of 0.005.
Thus, spurious motions are even smaller for this model.

\subsubsection{Convectively Unstable Atmospheres}

The method proposed here is able to reproduce convectively stable profiles with
Mach number fluctuations below 0.02. Thus, modeling convectively unstable
profiles should not be influenced by these velocities if the expected Mach
number of the convective motions is larger.

As an example of a convectively unstable model, we study \modelA{} that is
similar to \modelH{} studied in the previous section, but is reconstructed
using the density profile with the OPAL EOS (the same as in the \textsc{mesa} code in
this regime) and thus reproduces the convective behaviour of the original
$2M_\odot$ model. 
The thermal timescale of this stellar model is 3400\,yr, much longer
than the dynamic or sound crossing time (9.4\,d). It is also much longer than the
time covered by simulations of binary interactions using this giant, which is of
the order of a few years. Thus, the impact of radiative transfer should be
small, at least in the initial phase of the binary simulations.

The Mach number and density profiles are plotted in Fig.~\ref{fig:profileMrhoA}
together with those of the original \textsc{mesa} model. The density profile (lower panel)
reproduces the original profile quite well throughout the envelope and differs
only in the inner part, where it was cut off, and near the surface, where it
expands slightly. The difference becomes significant only below $\rho \approx
\SI{e-7}{\g\per\cubic\cm}$ and as the transition to the ambient medium is
gradual, the definition of a new radius would be ambiguous. Although convective
motions are present, the density profile changes only slightly near the surface
after the damping is disabled and is nearly stationary in the rest of
the envelope.

The Mach number (Fig.~\ref{fig:profileMrhoA}, upper panel) evolves from a state
similar to the non-convective model in the beginning (cf.
Fig.~\ref{fig:profileMrhoH}, upper panel) to a state with a Mach number between
0.1 and 0.2 throughout the envelope. The mean Mach number in this model is 0.12.
This indicates that convective motions develop and dominate the flow in the
envelope.
The stratification of the envelope is only marginally unstable, and the
convective motions are triggered by spurious velocity fluctuations.
Outside of the star, spurious Mach number fluctuations can be seen
similar to the non-convective model. Compared to the Mach number of convective
motions according to the mixing-length theory as computed by \textsc{mesa}, the Mach
numbers in the hydrodynamical model are much higher in most parts of the
envelope. The increase of the Mach number near the surface, however, is similar
in both the \textsc{mesa} model and our hydrodynamical model and is due to the decrease in
sound speed in the outer layers. The kinetic energy due to the convective
motions in this model is 1\% of the total energy. This value is similar for the
other convective models in Tab.~\ref{tab:relaxationruns} whereas for the
non-convective models, the kinetic energy is due to numerical fluctuations and
amounts to values between 0.01\% and 0.1\% of the total energy.

The Mach number distribution in the $x$-$y$ plane is compared in
Fig.~\ref{fig:comparisonmachnumber} for the non-convective \modelH{} (left
panel) and for the convective \modelA{} (right panel). Both
simulations show spurious fluctuations in the outer region that is
discretized in coarse cells.\footnote{This is because the outer region has a low
density and because the refinement criterion ensures similar cell masses; thus
low-density cells have large volumes. Because of the low density, the Mach
number fluctuations here can be neglected.} The non-convective model (left panel)
shows Mach numbers up to 0.05 only near the center and near the surface,
as also seen in the radial profile (Fig.~\ref{fig:profileMrhoH}, upper panel).
For the convective model (Fig.~\ref{fig:comparisonmachnumber}, right panel), a
complex flow pattern emerges after the damping is turned off.
It shows large plumes near the surface where the Mach number can reach values up
to 0.5 and features updrafts and downdrafts during its evolution. Thus, the
convectively unstable \modelA{} develops a convective pattern that can be
followed in the hydrodynamical simulation. The global convective
timescale for
this model is 58\,d (as estimated from integrating over
averaged convective velocities),
and the total time of the relaxation run is 94\,d (ten times the dynamical
timescale). Thus, the second half of the run, where no damping is applied,
covers roughly one convective timescale, so the development of
convection is expected during the run.
Although we cannot reproduce a steady state of the convective flow
field, the large-scale convective motions are recovered. These contain most of
the kinetic energy and thus potentially have the largest impact on the evolution
of the binary system.

The agreement in the hydrostatic equilibrium
is not expected to be perfect due to the convective motions in the
envelope.  The mean deviation for this model is 2.7\% and thus a factor of 3 larger
than in the non-convective case (cf.\ Tab.~\ref{tab:relaxationruns}), but still
on the per cent level. This indicates that the dynamic motions induced by
convection lead to changes in $\rho$ and increase the mismatch of terms in the
hydrostatic equilibrium, i.e., between pressure gradient and gravity.

Thus, convective envelopes of RG models can be reproduced for which a stationary
density structure results that follows the original model. Furthermore, the
model still approximately fulfills the hydrostatic equilibrium at the per cent
level and shows convective motions as expected in the envelope.

\subsubsection{Equation of State}

All hydrodynamical simulations of CE phases published so far use an ideal gas
EOS with constant $\gamma$ for the hydrodynamics
-- except for the SPH simulations by \citet{nandez2015a} and
\citet{nandez2016a}. Since stellar
evolution codes usually employ an EOS that accounts for the
ionization state of the plasma, the stellar profile has different thermodynamic
properties when using an ideal gas EOS\@. As shown in
Sec.~\ref{sec:reconstruction} (cf.\ Fig.~\ref{fig:comprgideal}), this leads to
changes in the convective stability. The thermodynamic structure and the
convective behaviour can only be reproduced using the same (or a very similar)
EOS\@.

\ModelJ{} of our simulation suite was set up by reconstructing the stellar
profile of the $2M_\odot$ RG using the density profile and an ideal gas EOS with
$\gamma = 5/3$. According to the stellar evolution code (that uses the OPAL
EOS), this model has a deep convective envelope. However, the reconstruction
with the ideal gas EOS leads to a convectively stable stratification (cf.
Fig.~\ref{fig:comprgideal}). This is reproduced by \modelJ{}; it does not
develop convective motions: the mean Mach number remains as low as $\num{6e-3}$
and the mean error in the hydrostatic equilibrium is $1.3\%$ (see
Tab.~\ref{tab:relaxationruns}). In this respect, the model is very similar to
the artificially stabilized \modelG{} at the same resolution. Thus,
mapping the density and pressure profile of this model from the stellar
evolution code to the hydrodynamics code and using an ideal EOS results in a
suppression of convective motions that should be present in the convective
envelope. 

To recover the convective properties of the stellar model, one can reconstruct
$\nabla - \nabla_\mathrm{ad}$, i.e., the difference to the adiabatic gradient
(see Sec.~\ref{sec:reconstruction}). Setting this to zero for the reconstruction
yields an adiabatic stratification that is convectively unstable.  
Although the stratification is only marginally unstable, spurious
velocity fluctuations due to mismatches in the discretization of the hydrostatic
equilibrium trigger the onset of convective motions in the envelope.
Since this reconstruction method
leads to a shallower temperature gradient, a smaller envelope results with less
mass (see Fig.~\ref{fig:comprgideal} and \modelK{} in
Tab.~\ref{tab:relaxationruns}). The relaxation run for \modelK{} shows
convective motions, as expected for a convectively unstable model: the mean Mach
number is 0.045 and the mean hydrostatic equilibrium error 1.6\%. The mean Mach
number is smaller than for the models reconstructed using the OPAL EOS which
have a mean Mach number of about 0.12 (\modelA{} and \modelB{}, see
Tab.~\ref{tab:relaxationruns}). This is probably due to the different
stratification.

Hence, the expected behaviour of initial stellar profiles can be reproduced also
using an ideal EOS\@.  The convective envelope of the giant, however, is
expected to be convectively stable (as opposed to the original \textsc{mesa}
model) when reconstructing the profile with an ideal EOS\@.  Thus, to generate
stellar models for the hydrodynamical simulations that show the correct
convective behaviour, the same equation of state has to be used as in the
stellar evolution code.

\subsubsection{Numerical Parameters}
\label{sec:numericalparameters}

The construction of the initial models is influenced by the choice of the
parameters resolution, core cutoff radius, and grid type.

To study the convergence behaviour of the simulations, the convectively stable
model created by reconstructing the profile of $\nabla$ with a shift of
0.02 was run in two resolutions: with $\num{5e5}$ cells (\modelG{}) and with
$\num{2e6}$ cells (\modelH{}). For these models, the mean Mach number
fluctuations are similar ($\num{9.6e-3}$ vs.  $\num{9.1e-3}$). The mean error in
the hydrostatic equilibrium decreases by a factor of 1.6 which corresponds to
the increase in resolution by a factor of 4 in number of cells which roughly
means a factor of $4^{1/3}\approx 1.6$ in average spatial resolution; hence, the
error decreases to first order with the mean linear cell size.
The convectively unstable model reconstructed using the density profile and the
OPAL EOS was run in three resolutions (see Tab.~\ref{tab:relaxationruns}): with
$\num{e5}$ cells (\modelC{}), with $\num{5e5}$ cells (\modelB{}), and with
$\num{2e6}$ cells (\modelA{}).  At the lowest resolution, the mean Mach number
(0.075) was smaller than for the other resolutions (0.12 for both); the errors
in the hydrostatic equilibrium are similar and probably dominated by the
presence of convective motions. Since the two highest resolutions show very
similar properties, we conclude that the resolution of the convective motions is
converged when using more than $\num{5e5}$ cells.  

The properties of simulations using different core cutoff radii are quite
similar (compare \modelB{}, \modelD{}, \modelE{}), although for a large
radius (10\%, \modelE{}), the mean Mach number is slightly smaller. Decreasing
the core cutoff radius is a goal for CE simulations because in this way, a
larger fraction of the envelope can be reproduced on the grid. For convectively
stable models, the model with a smaller core cutoff radius (\modelF{},
core is 2\% of radius) shows a larger value of the mean Mach number by a factor
of 2 than the \modelG{} (core is 5\% of radius).  Here, probably a higher
resolution is needed to resolve the hydrostatic equilibrium better where the
largest pressure gradients occur, namely at the boundary between core and
envelope.

All simulations presented so far used as initial grid configuration spherical
shells with cells distributed on these shells according to the HEALPix
distribution (cf. Sec.~\ref{sec:grids}). Simulations similar to 
\modelB{} have been also carried out for different initial grids: 
\modelL{} using a cubic grid, \modelM{} using a cubic adaptive grid, and
\modelN{} using a cubic nested grid (see Tab.~\ref{tab:relaxationruns}).
However, because a moving mesh is employed with an adaptive refinement ensuring
similar cell masses, the difference between these simulations is small and the
influence of the initial grid is negligible. This is, however, not true for
static meshes: tests on static meshes show that the mass-adaptive grids result
in better equilibria since they reflect the symmetry and allow for high
spatial resolution in the center.

\subsection{AGB models}

\begin{table*}
  %\scriptsize
  %\footnotesize
  %\small
  \centering
  \caption{\textbf{Overview of AGB Relaxation Runs.} The table shows data for 2
different simulations for the initial $0.67M_\odot$ AGB model. The columns display a
model letter,
the equation of state,  the
radius and gas mass of the reconstructed profile in solar units, the mean Mach
number at the end of the relaxation run and the mean relative difference in the
hydrostatic equilibrium. The mean values are computed using the cell mass as
weight. Both simulations are reconstructed using the density and mapped on a
HEALPix grid with a resolution of $\num{5e5}$ cells. The core mass in both
simulations is $0.54M_\odot$, the total mass $0.68M_\odot$ ($0.67M_\odot$ in the
\textsc{mesa} model).}
\label{tab:relaxationrunsagb}
\vspace{2mm}
  \begin{tabular}{lcS[table-format=3]S[table-format=1.3]S[table-format=1.1e-1,table-auto-round]S[table-format=1.1e-1,table-auto-round]}
  %\toprule
  \hline\hline
                     Model &   EOS & 
  {$\frac{R}{R_\odot}$} & {$\frac{M_\mathrm{env}}{M_\odot}$} & {$\overline{\mathcal{M}}$} &
  {$\overline{\frac{|\rho\vec{g} - \nabla p|}{\max(|\rho\vec{g}|,|\nabla p|)}}$} \\
  %\midrule
  \hline
O &     OPAL & 166 & 0.137 & 2.09e-01 & 8.18e-02 \\ %                AGB-OPAL & 
P &    Ideal & 166 & 0.137 & 1.78e-02 & 1.70e-02 \\ %               AGB-Ideal & 
  %\bottomrule
  \hline
  \end{tabular}
\end{table*}

\newcommand{\modelO}[0]{model~O}
\newcommand{\modelP}[0]{model~P}
\newcommand{\ModelO}[0]{Model~O}
\newcommand{\ModelP}[0]{Model~P}

\begin{figure*}[tbp]
  \centering
    \includegraphics[width=0.66\textwidth]{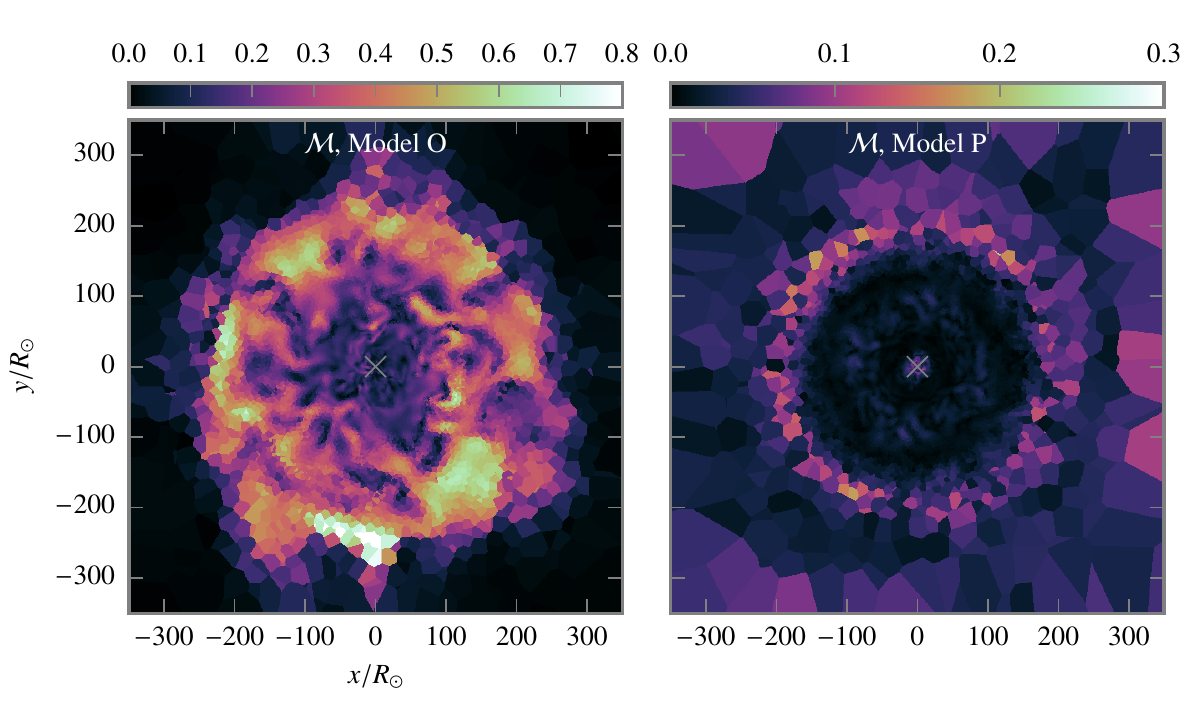}
  \caption{{Comparison of Mach number distributions for
      \modelO{} (left panel) and \modelP{} (right panel).} \ModelO{} should be
      convectively unstable, \modelP{} convectively stable.}
\label{fig:comparisonmachnumberagb}
\end{figure*}

\begin{figure*}[tbp]
  \centering
    \includegraphics[width=0.66\textwidth]{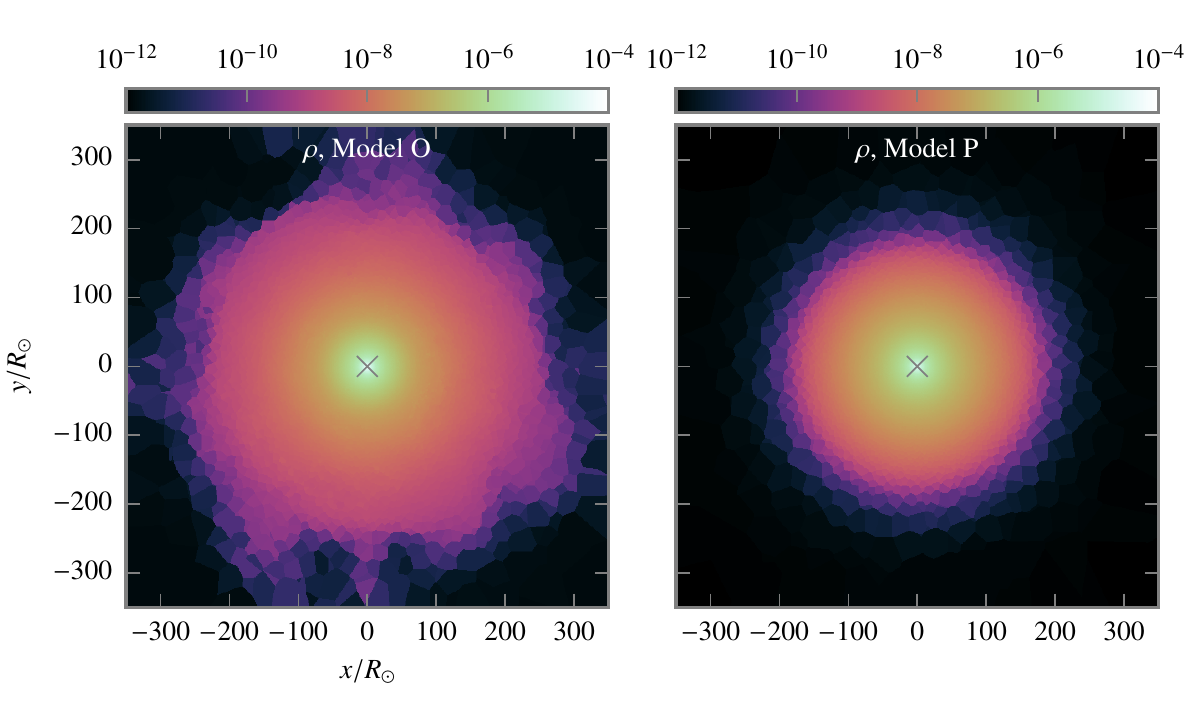}
  \caption{{Comparison of density distributions for
      \modelO{} (left panel) and \modelP{} (right panel).} \ModelO{} should be
      convectively unstable, \modelP{} convectively stable.}
\label{fig:comparisondensityagb}
\end{figure*}

In addition to the $2M_\odot$ RG model presented in the previous section, a
$0.68M_\odot$ AGB star (from a ZAMS model of $1M_\odot$) is used as a
second example. Its luminosity is $2400L_\odot$, the dynamical timescale is
75\,d, and the thermal timescale is 27\,yr. The stellar model
was set up using the
OPAL EOS and the ideal EOS (Tab.~\ref{tab:relaxationrunsagb}). Both models were
reconstructed using the density and mapped to a HEALPix grid with $\num{5e5}$
cells. Similar to the $2M_\odot$ RG model, the AGB model using the ideal EOS is
expected to be convectively stable in most parts of the envelope (except near
the surface and in a zone at a radius of about $80R_\odot$), whereas the model
using the OPAL EOS should be convectively unstable throughout the envelope.

This behaviour is indeed reproduced by our simulations: the mean Mach number in
the convectively unstable \modelO{} is 0.2, whereas it is about 0.02 for
the convectively stable \modelP{}. Moreover, the distribution of the Mach
number in the $x$-$y$ plane (Fig.~\ref{fig:comparisonmachnumberagb}) shows
convective motions and large plumes with Mach numbers up to 0.8 for 
\modelO{} (left panel). \ModelP{} (right panel) shows only small Mach
numbers in the envelope that are located in the central region, near the
surface, and in a shell at approximately the radius ($\approx 80R_\odot$) that
is expected to be convectively unstable.

\ModelP{} also shows a spherical density distribution following the
original \textsc{mesa} model (Fig.~\ref{fig:comparisondensityagb}, right panel) and the
mean error of the hydrostatic equilibrium is about 2\%. Hence, this model is
still in a good hydrostatic equilibrium.  \ModelO{}, however, shows a very
dynamic behaviour: the convective cells are comparable to the size of the star
and lead to a non-spherical density distribution that is also more extended
compared to the original \textsc{mesa} model (Fig.~\ref{fig:comparisondensityagb}, left
panel). 

The total runtime of the relaxation covers 750\,d, i.e., roughly
2\,yr. This is only one order of magnitude smaller than the thermal timescale;
thus, radiative transfer may be important for binary simulations that cover
longer time spans.

\section{Discussion}
\label{sec:discussion}

\subsection{Comparison to other setups used in simulations of the CE phase}

Hydrodynamical simulations of the CE phase with grid codes have employed two
strategies of dealing with the wide range of scales in giant profiles
up to now: either the core is replaced by a point mass that interacts only
gravitationally, similar to the method proposed in this paper 
\citep[e.g.][]{sandquist1998a,sandquist2000a,passy2012a,staff2016b,staff2016a,iaconi2016a},
or the core is represented by a cloud of particles for which the density is
mapped to the grid for computing gravity 
\citep[cloud-in-cell method,][]{ricker2008a,ricker2012a}. 
Both types of simulations did not employ a continuation of the profile below a
certain radius, but simply mapped the one-dimensional profile to the grid, with
the center of the star being in the center of a grid cell. This implies cutting
the profile of the giant at a certain radius that depends on the resolution;
moreover, the central part of the mapped profile is not in hydrostatic
equilibrium, leading to velocity fluctuations. Hence, usually a strong damping
is employed, reducing the velocities by about a factor of two each timestep.
This implies damping on a timescale of the order of the timestep.

In contrast, the method proposed in this paper leads to a one-dimensional
profile that can be mapped to the hydrodynamical grid independent of its
resolution. Furthermore, the damping timescale can be chosen to be a tenth of
the dynamical timescale (i.e., much larger than a timestep, see
Eq.~\ref{eq:relaxationtime}), which damps sound waves on a physical timescale,
allowing the profile to settle to an equilibrium. The damping can be made weaker
with this method because the one-dimensional profile incorporates the
gravitational force of the point mass and leads to a stable hydrostatic
equilibrium after it is mapped to the grid. Thus, only slight distortions are
generated, leading to small velocity fluctuations that can easily be damped
away.

Moreover, grid simulations have up to now employed an ideal gas EOS that usually
leads to convectively stable envelopes although the envelope of the giant is
convectively unstable in the stellar evolution code (cf.\ 
\modelJ{} in Tab.~\ref{tab:relaxationruns}). Using a similar EOS in the
hydrodynamical simulations as in the stellar evolution code, the convective
behaviour of the model can be reproduced in our simulations. The reconstruction
method using the temperature gradient (see Eq.~\ref{eq:integrategivennabla}) can
also be used to specifically generate models that are convectively stable or
unstable, although this may lead to differences in the mechanical structure.
If using a different EOS in the hydrodynamics code than in the stellar
evolution code (e.g., an ideal gas EOS), either the mechanical profile of the
star can be reproduced (i.e., density and pressure profiles) or its convective
behaviour, but not both. In previous simulations, only the mechanical
structure was recovered. With the method proposed here, however, it is possible
to choose which properties of the stellar model are reproduced.

\subsection{AGB models}

The AGB models show rather violent behaviour with high Mach numbers up to 0.8
and with large convective cells, resulting in a dynamic structure that is not
spherically symmetric.  This behaviour is similar to what is seen in the
radiation hydrodynamical simulations by \citet{freytag2008a} and
\citet{freytag2012a}. Their code uses a Cartesian grid, fixes the gravitational
field, and includes non-local radiative transport. Moreover, energy is inserted
in the interior with a magnitude corresponding to the core luminosity that
drives convection. The simulations display a complex dynamical behaviour with
large convective cells and shocks in the envelope. Although we do not include
radiative transport, the convective motions in the envelope look qualitatively
similar in our models. These, however, are tailored to be used
as initial models for binary interactions that include self-gravity.

\section{Conclusions}
\label{sec:conclusions}

To study hydrodynamical interactions involving giant stars -- such as CE phases --
initial conditions for the hydrodynamical simulations have to be generated from
the stellar evolution model. In this paper, a new method is presented to
reliably create approximate models of giants that can be used for this purpose.

The profiles of giants are approximated by replacing the core with a point mass
to limit the range in spatial and temporal scales such that hydrodynamical
simulations are computationally feasible. The profile is continued to the
stellar center by solving a modified Lane--Emden equation, a method that does
not depend on the resolution of the grid in the hydrodynamical simulation. When
using a different EOS in the hydrodynamical simulation than that in the stellar
evolution code, the thermodynamic properties change. An improved reconstruction
method is proposed that allows to recover either the mechanical properties
(i.e., the density and pressure profiles) or the convective behaviour of the
envelope. When using the same EOS, both methods yield the same profile as in the
stellar evolution calculation.

These reconstructed profiles of giant envelopes are then mapped onto
an unstructured grid for hydrodynamical simulations. To reach a stable model,
a relaxation procedure is utilized: spurious velocity fluctuations are damped on
a timescale related to the dynamical timescale of the giant. Several criteria
are checked to assess the stability of the envelope: the profiles of density and
pressure should remain similar to the original ones; the hydrostatic equilibrium
should be maintained; the Mach number should be small for a convectively stable
envelope, and a steady state should develop for a convectively unstable
envelope; no pulsations should be present. It is shown that these criteria are
fulfilled for the examples of a $2M_\odot$ RGB and a $0.68 M_\odot$ AGB star;
extending these simulations to other masses and evolutionary stages is
straightforward.  The examples show that convectively stable envelopes can be
reproduced reliably, where the convective stability can be controlled by the
reconstruction method.  Convectively unstable envelopes develop fluid motions
with profiles similar to the stellar evolution models.

The method proposed here improves on previous hydrodynamics studies
\citep{sandquist1998a,sandquist2000a,ricker2008a,ricker2012a,passy2012a,staff2016b,staff2016a,iaconi2016a},
by generating approximate profiles that do not depend on the grid resolution and
by allowing to control the convective behaviour of the envelope by the
reconstruction method. Also, the relaxation procedure is shown to generate giant
envelopes that remain stable for several dynamical timescales.

Stable giant envelopes as modeled here can be used as initial
conditions for hydrodynamical simulations of the CE phase
\citep{ohlmann2016a,ohlmann2016b},
but might also be used to study mass transfer in binary systems or other
dynamical phenomena involving giant envelopes.

\begin{acknowledgements}
The work of STO was supported by Studienstiftung des deutschen Volkes
and by the graduate school GRK 1147 at University W\"urzburg.
STO and FKR acknowledge support by the DAAD/Go8 German-Australian exchange
program. RP and VS acknowledge support by the European Research Council under
ERC-StG grant EXAGAL-308037. 
STO, FKR, RP, and VS were supported by the Klaus Tschira Foundation.
Parts of this work were performed on the computational resource ForHLR~I
funded by the Ministry of Science, Research and the Arts Baden-Württemberg and
DFG (``Deutsche Forschungsgemeinschaft'').
For data processing and plotting, NumPy and SciPy
\citep{oliphant2007a}, IPython \citep{perez2007a}, and Matplotlib
\citep{hunter2007a} were used.
We thank the anonymous referee for helpful comments.
\end{acknowledgements}

\bibliographystyle{aa} % style aa.bst
\bibliography{ref}

\begin{appendix}
\section{Derivation of Mach number fluctuations}
\label{sec:derivation}

The magnitude of the velocity fluctuations introduced by the spatial
discretization can be computed for the \textsc{arepo} scheme as follows. For an
overview of numerical hydrodynamics and an in-depth explanation of Godunov
schemes and Riemann solvers, see the book by \citet{toro2009a}.

The initial condition is chosen as an isothermal atmosphere given in $z$ direction as $p(z)
= p_0 \exp(-z/H)$ and $\rho(z) = \rho_0 \exp(-z/H)$, with $H$ denoting the
pressure scale height. We assume an ideal equation of state. The gravitational
acceleration is constant and given by $g = \frac{p_0}{H\rho_0}$. The speed of
sound is also constant and given by $c_\mathrm{s} = \sqrt{\gamma P/\rho}$. The
grid is assumed to be equidistant with grid spacing $\delta$. We use the HLLC
solver \citep{toro1994a} to compute the Riemann problems at the cell interfaces. 

For this setup, the update of the conserved
quantities $\vec{U}_i$ for cell $i$ reads for a simple forward Euler time step
(from time $t^n$ to $t^{n+1}= t^n + \Delta t$):
\begin{equation}
  \vec{U}_i^{(n+1)} = \vec{U}_i^{(n)} - \frac{\Delta t}{\delta}
  \left(\vec{F}_{i+\frac{1}{2}} - \vec{F}_{i-\frac{1}{2}}\right) 
  + \Delta t \vec{S}_i,
  \label{eq:updatescheme}
\end{equation}
where $\vec{F}_{i\pm \frac{1}{2}}$ denotes the flux over the left (right) cell
boundary and $\vec{S}_i = (0, \rho_i g, \rho_i g v_i)^\intercal$ is the
gravitational source term ($\rho_i$: density, $v_i$: velocity). 

For the initial conditions given here, the velocity is zero and for each Riemann
problem, $p_\mathrm{L} > p_\mathrm{R}$ (L: left interface, R: right interface).
Thus, the fluxes are computed for the subsonic L$^*$ region of the HLLC solver
\citep{toro1994a} as
\begin{equation}
  \vec{F}_\mathrm{L}^* = \vec{F}_\mathrm{L} + S_\mathrm{L}
  \left(\vec{U}_\mathrm{L}^* - \vec{U}_\mathrm{L}\right),
  \label{eq:fluxlstar}
\end{equation}
where $S_\mathrm{L}$ is the left wave speed and $\vec{U}_\mathrm{L}^*$ the
solution of the Riemann problem in the L$^*$ region. $\vec{F}_\mathrm{L}$ is the
Euler flux computed with the values at the left interface. Plugging in the
solution of the Riemann problem as given in \citet{toro1994a} for zero
velocities ($v_\mathrm{L} = v_\mathrm{R} = 0$) yields for the momentum flux
\begin{equation}
  p_\mathrm{L} + \frac{S_\mathrm{L}^2 S_\mathrm{M}}{S_\mathrm{L} +
  S_\mathrm{M}} \rho_\mathrm{L},
  \label{eq:momflux}
\end{equation}
where $S_\mathrm{M}$ denotes the velocity of the contact discontinuity.
This flux can now be combined with Eq.~\ref{eq:updatescheme} to compute the
update of the momentum,
\begin{multline}
  (\rho v)_i^{(n+1)} = - \frac{\Delta t}{\delta} \left( 
  p_{i+\frac{1}{2}}^\mathrm{L} - p_{i-\frac{1}{2}}^\mathrm{L} \right.\\\left.
  + \left( \frac{S_\mathrm{L}^2 S_\mathrm{M}}{S_\mathrm{L} +
  S_\mathrm{M}} \rho_\mathrm{L} \right)_{i+\frac{1}{2}} 
  - \left( \frac{S_\mathrm{L}^2 S_\mathrm{M}}{S_\mathrm{L} +
  S_\mathrm{M}} \rho_\mathrm{L} \right)_{i-\frac{1}{2}}
  \right)
  + \Delta t \rho_i g,
  \label{eq:updatemomentum}
\end{multline}
where $v^{(n)}=0$ is assumed. To obtain the magnitude of velocity fluctuations
caused by the spatial discretization, a reconstruction has to be chosen. For
constant reconstruction, $p_{i+\frac{1}{2}}^\mathrm{L}=p_i$ holds, whereas for
linear reconstruction a central finite difference for the derivative is employed
which yields
\begin{equation}
  p_{i+\frac{1}{2}}^\mathrm{L} = p_i + \frac{p_{i+1} - p_{i-1}}{2\delta}
  \frac{\delta}{2} = p_i + \frac{1}{4} \left(p_{i+1} - p_{i-1}\right).
  \label{eq:linreconstruction}
\end{equation}

After inserting the reconstruction into Eq.~(\ref{eq:updatemomentum}),
pressure and density are expanded into a Taylor series around cell $i$ and only
the highest order terms are retained. For this expansion, we assume that the
value $p_i$ represents the pressure at $x_i$ and not the average over cell
$i$, which is the usual interpretation, because we use the same procedure in the
setup of our simulations\footnote{If $p_i$ is taken as the average over
cell $i$, the result remains the same for constant reconstruction. For linear
reconstruction, the constant in Eq.~(\ref{eq:machfluctuations}) changes from
$1/12$ to $7/72$, i.e., by roughly 20\%. The order of magnitude and the scaling
with $\delta/H$ stays the same.}.  For better comparison between different
initial values, the resulting Mach number $M = v / c_\mathrm{s}$ is computed.
The first order term in the expansion cancels out the gravitational source
term, but for constant reconstruction, Mach number fluctuations are of the order
of $(\delta/H)^2$. For linear reconstruction, this order vanishes as well and
we obtain Eq.~(\ref{eq:machfluctuations}).

A similar derivation was done by \citet[Sec.~3.1]{zingale2002a}, but for a PPM
scheme in the context of mapping hydrostatic profiles into the AMR code
FLASH\@. For their scheme, they find a coefficient of $5/24$ and the same order
of the error, similar to our findings.

\end{appendix}

\end{document}